\documentclass{article}
\usepackage{PRIMEarxiv}

\usepackage{hyperref}
\usepackage{color}
\usepackage{pifont}
\usepackage[many]{tcolorbox} 
\usepackage{paralist}
\usepackage{subfigure}
\usepackage{cleveref}
\usepackage{microtype}
\usepackage{fancyhdr}       
\usepackage{graphicx}      
\usepackage{ragged2e}
\usepackage[numbers]{natbib}

\usepackage{enumitem}
\usepackage[dvipsnames]{xcolor}
\usepackage{multirow}
\usepackage{colortbl}
\usepackage{longtable}
\usepackage{afterpage}

\definecolor{objmain}{HTML}{5989cf}

\newcounter{sbseStrength}
\newcounter{sbseWeakness}
\newcounter{fmStrength}
\newcounter{fmWeakness}

\newcommand{\sbsestrength}{%
\refstepcounter{sbseStrength}%
\revision{\textbf{SB-S\thesbseStrength}}%
\label{SB-S\thesbseStrength}%
}
\newcommand{\sbseweakness}{%
\refstepcounter{sbseWeakness}%
\revision{\textbf{SB-W\thesbseWeakness}}%
\label{SB-W\thesbseWeakness}%
}

\newcommand{\fmstrength}{%
\refstepcounter{fmStrength}%
\revision{\textbf{FM-S\thefmStrength}}%
\label{FM-S\thefmStrength}%
}
\newcommand{\fmweakness}{%
\refstepcounter{fmWeakness}%
\revision{\textbf{FM-W\thefmWeakness}}%
\label{FM-W\thefmWeakness}%
}

\newcommand{\fullref}[1]{\hyperref[#1]{\textcolor{blue}{\textbf{#1}}}}

\newcommand{\revision}[1]{\textcolor{black}{#1}}

\title{
Search-Based Software Engineering and AI Foundation Models: Current Landscape and Future Roadmap
}

\author{
  Hassan Sartaj \\
  Simula Research Laboratory, Oslo, Norway\\
  University of Central Punjab, Lahore, Pakistan\\
  \texttt{hassan@simula.no,hassan.sartaj@ucp.edu.pk} \\
  \And
  Shaukat Ali \\
  Simula Research Laboratory \\
  Oslo, Norway\\
  \texttt{shaukat@simula.no} \\
  \And
  Paolo Arcaini \\
  National Institute of Informatics \\
  Tokyo, Japan\\
  \texttt{arcaini@nii.ac.jp} \\
  \And
  Andrea Arcuri \\
  Kristiania University of Applied Sciences and Oslo Metropolitan University \\
  Oslo, Norway\\
  \texttt{andrea.arcuri@kristiania.no} \\
}

\begin{document}
\maketitle

\begin{abstract}
Search-based software engineering (SBSE), which integrates metaheuristic search techniques with software engineering, has been an active area of research for about 25 years. It has been applied to solve numerous problems across the entire software engineering lifecycle and has demonstrated its versatility in multiple domains. With recent advances in Artificial Intelligence (AI), particularly the emergence of foundation models (FMs) such as large language models (LLMs), the evolution of SBSE alongside these models remains undetermined. In this window of opportunity, we present a research roadmap that articulates the current landscape of SBSE in relation to FMs, identifies open challenges, and outlines potential research directions to advance SBSE through its synergy with FMs. 
Specifically, we analyze three core aspects: utilizing FMs to enhance SBSE, applying SBSE to advance FMs, and exploring the integration of SBSE and FMs. Furthermore, we present a forward-thinking perspective that envisions the future of SBSE in the era of FMs, highlighting promising research opportunities to address challenges in emerging domains.
\end{abstract}


\section{Introduction}
Search-Based Software Engineering (SBSE), formally introduced as a field by~\citet{harman2001search} in 2001, has gained prominence over the past quarter century, although the use of search algorithms to address software engineering (SE) problems dates back to earlier work, such as test data generation by \citet{miller1976automatic} in 1976. 
Over the years, it has rapidly evolved to address emerging and complex SE problems. 
It has been successfully applied throughout the SE lifecycle, including requirements engineering, software design, software development, software testing, deployment, and maintenance~\cite{harman2012search}.

SBSE applies artificial intelligence (AI) techniques, including evolutionary computation and metaheuristic search, to effectively solve SE problems. 
This involves reformulating SE problems as metaheuristic search problems by creating solution representation, defining fitness functions, and selecting search operators~\cite{harman2001search}.
The problem formulated in this way is then solved by applying search and optimization techniques to identify optimal solutions. 
These techniques include single-objective, multi-objective, and many-objective search algorithms, each designed for different problem types and levels of complexity~\cite{harman2009theoretical, harman2012search}.

Recently, foundation models (FMs)---a term introduced by~\citet{bommasani2021opportunities}---have brought a major transformation, attracting widespread interest across research communities, academia, and industry~\cite{li2025sefms,llmsForSEZhang2023arxiv}. 
Although the term ``foundation models'' broadly refers to large-scale pre-trained models, it is most commonly associated with large transformer-based architectures, such as GPT, BERT, and CLIP~\cite{bommasani2021opportunities}. 
These models are trained on vast amounts of data and are capable of performing a wide range of analytical tasks. 
For specific application domains, these models can be adapted using techniques such as fine-tuning and prompt engineering. 
Depending on their modalities, architecture, and application domains, FMs can be categorized as follows: Large Language Models (LLMs) like GPT for textual content; Vision Language Models (VLMs) like ResNet for image and video data; Speech and Audio Models (SAMs) like WaveNet for tasks related to speech recognition, synthesis, and audio analysis; and Multimodal Models (MMs) like CLIP that integrate multiple types of data, such as text and images, to enable content generation and comprehension across different modalities~\cite{LMMsDawn2023}. 
Their ability to generalize knowledge across multiple domains and adapt to diverse applications has enabled new AI-driven innovations, making them a focal point of exploration. 
As a result, researchers are actively exploring their potential, academic institutions are integrating them into curricula, and industries are leveraging them to enhance automation, decision-making, and user experiences~\cite{sun2025survey}.

Considering rapid AI transformations driven by FMs, in this paper, we present a research roadmap exploring the \revision{synergy} between SBSE and FMs.
We first analyze the current research landscape to examine the directions in which SBSE and FMs are being integrated and the types of SE problems they address. 
From this analysis, we identify open challenges and opportunities for developing novel techniques that combine SBSE and FMs to solve complex domain-specific and SE problems. 
We also highlight the challenges of conducting empirical evaluations of such techniques, which require significant attention from the research community. 
Finally, we outline a 2030 research horizon, using McLuhan's Tetrad~\cite{mcluhan1977laws}---a framework designed to analyze the effects of new technologies---to illustrate how FMs, as disruptive technology, can shape the future of SBSE. 
Furthermore, we present research opportunities for potential synergy between SBSE and FMs to address problems in emerging domains.

\section{Roadmap Overview}\label{sec:overview}
\begin{figure}
\centering
\includegraphics[width=1\linewidth]{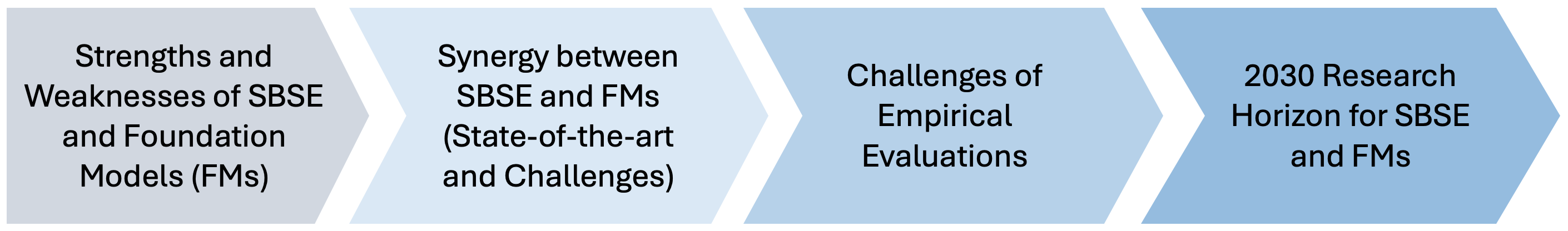}
\caption{Roadmap overview showing the discussion flow: strengths and weaknesses (\Cref{sec:pros&cons}), SBSE–FM synergies (\Cref{sec:RootFMs4SBSE,sec:RootSBSE4FMs,sec:SBSE&FMs}), empirical evaluation challenges (\Cref{sec:empiricalchallenges}), and the 2030 research horizon (\Cref{sec:horizon}).}
\label{fig:outline}
\end{figure}

\begin{figure}[t]
\centering
\includegraphics[width=0.8\linewidth]{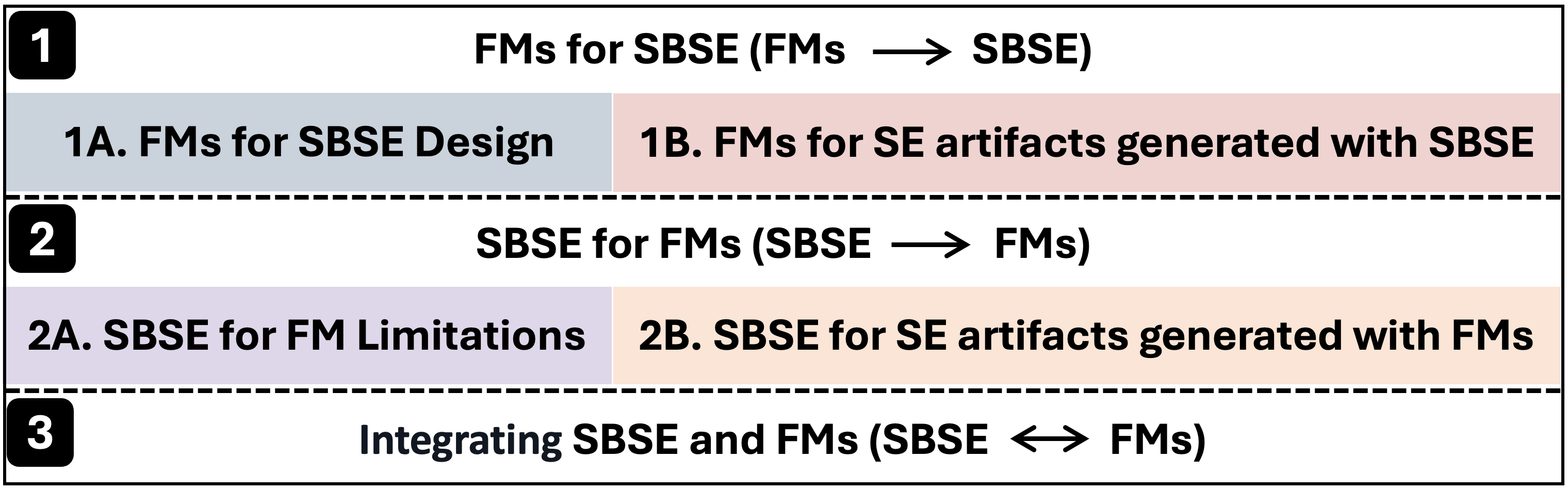}
\caption{Key aspects of the potential synergy between SBSE and FMs. The abbreviations used are: FMs (Foundation Models), SBSE (Search-Based Software Engineering), and SE (Software Engineering).}
\label{fig:overallsynergy}
\end{figure}

\Cref{fig:outline} illustrates the overall structure of the roadmap. 
We begin by describing the strengths and weaknesses of SBSE and FMs (\Cref{sec:pros&cons}). 
Following this, we explore the potential synergy between SBSE and FMs (as shown in~\Cref{fig:overallsynergy}), analyzing how the strengths of one can complement the other to advance SE (\Cref{sec:RootFMs4SBSE,sec:RootSBSE4FMs,sec:SBSE&FMs}).
Subsequently, we discuss the challenges associated with conducting empirical evaluations using techniques that integrate SBSE and FMs (\Cref{sec:empiricalchallenges}). 
Finally, we outline the research horizon for 2030, including open research challenges and opportunities ahead (\Cref{sec:horizon}).
\revision{
In the following, we describe the methodology followed to develop the roadmap and present the structure of the proposed roadmap. 
}

\begin{figure}[t]
\centering
\includegraphics[width=0.8\linewidth]{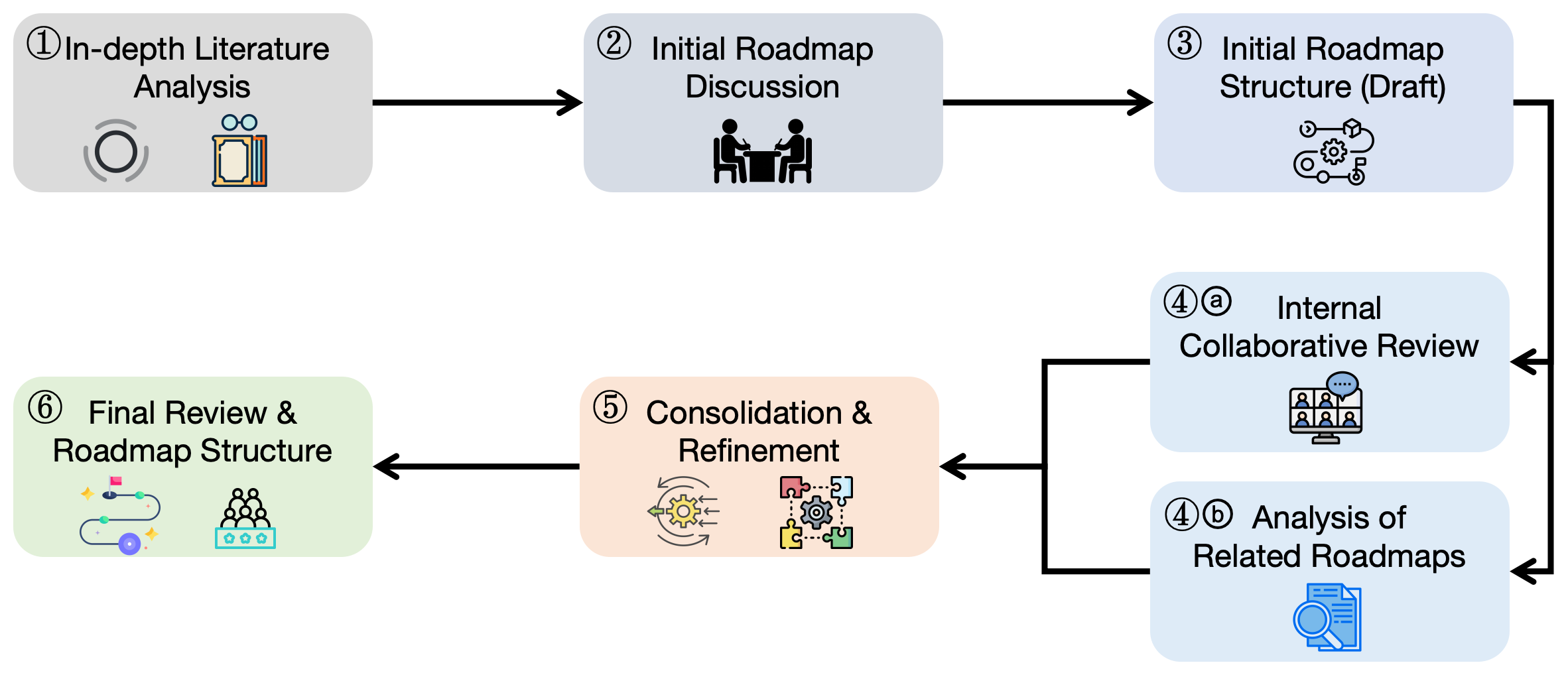}
\caption{\revision{Overview of the methodology used to develop the proposed roadmap.}}
\label{fig:mapmethod}
\end{figure}

\paragraph{\revision{Roadmap Development Methodology.}}

\revision{
The proposed roadmap was developed through an iterative and collaborative process involving literature analysis, expert discussions, and successive refinements, as illustrated in~\Cref{fig:mapmethod}. 
First, the first author reviewed the recent literature on SBSE and foundation models to identify research directions and emerging trends. 
Based on this analysis, the first and second authors held two in-person meetings (approximately one hour each) to discuss potential ways of organizing the roadmap and to identify its overall structure. 
Following these discussions, the first author prepared an initial roadmap structure. 
This initial version was subsequently reviewed during an online meeting with the third and fourth authors, where the proposed structure was critically evaluated. 
During this discussion, we also analyzed related roadmaps and identified the roadmap by~\citet{wu2024evolutionary} as the most relevant, which we used as inspiration to refine our roadmap structure.
Throughout these discussions, we followed a consensus-oriented process inspired by the Delphi method~\cite{linstone1975delphi} and Nominal Group Technique~\cite{harvey2012nominal}, enabling iterative refinement through structured discussion and expert feedback. 
In the subsequent step, we consolidated the feedback from all authors and insights drawn from the literature to refine the roadmap structure. 
Finally, the roadmap structure underwent a final review by all authors to ensure consistency, completeness, and alignment with the current state of research before identifying the research opportunities presented in this paper. 
}

\paragraph{\revision{Roadmap Structure.}}
\revision{
During the roadmap development process, we observed three emerging directions in research exploring the synergy between SBSE and FMs, as shown in~\Cref{fig:overallsynergy}. 
First (\ding{182} in the figure), FMs are being used to advance SBSE by:
\begin{inparaenum}[(a)]
\item supporting the design of SBSE techniques, e.g., fitness function definition, solution encoding, search space definition, and SBSE implementation generation; and
\item improving the software engineering artifacts generated using SBSE, e.g., artifact interpretation and refinement, Pareto analysis, and artifact validation. 
\end{inparaenum}
Second (\ding{183} in the figure), SBSE is being applied to FMs by:
\begin{inparaenum}[(a)]
\item addressing FM limitations, e.g., hallucination mitigation, uncertainty management, fine-tuning, and prompt optimization; and
\item adapting well-established SBSE techniques to optimize software engineering artifacts generated using FMs, e.g., code improvement, testing, model optimization, configuration optimization, and automated debugging and repair.
\end{inparaenum}
Third (\ding{184} in the figure), FMs are increasingly becoming an integral part of the search and optimization process within SBSE, e.g., by guiding search algorithms, fitness evaluation, and search operators, as well as integrating FMs and SBSE to solve software engineering problems. 
Accordingly, our roadmap focuses on the following research questions: 
}
\begin{inparaenum}[(1)]
\item how FMs can be leveraged to enhance SBSE (\Cref{sec:RootFMs4SBSE}), exploring (1A) FMs for SBSE design and (1B) their role in improving SBSE applied to SE problems, 
\item how SBSE can be employed to advance FMs (\Cref{sec:RootSBSE4FMs}), focusing on (2A) addressing FM limitations and (2B) adapting well-studied SBSE practices to FMs customized for particular SE activities, 
and
\item how SBSE and FMs operate together, \revision{where FMs become an integral part of the SBSE and lead to integrated solutions to address SE problems} (\Cref{sec:SBSE&FMs}).
\end{inparaenum}
For each aspect, we provide a review of current state-of-the-art research, followed by a roadmap with challenges and research opportunities.

\section{Strengths and Weaknesses of SBSE and FMs}\label{sec:pros&cons}
In this section, we begin by exploring the advantages and limitations of SBSE, followed by an analysis of the benefits and downsides of FMs. 
To highlight the strengths and weaknesses of SBSE and FMs, and connect them to subsequent sections, we use labels such as \textbf{SB-S} (SBSE Strength), \textbf{SB-W} (SBSE Weakness), \textbf{FM-S} (FM Strength), and \textbf{FM-W} (FM Weakness).

\subsection{Advantages and Limitations of SBSE}\label{subsec:sbse}

In SBSE~\cite{harman2001search}, SE problems are cast into either \emph{search} or \emph{optimization} problems.
Then, any kind of search or optimization algorithm can be applied to attempt to solve those SE problems.

On the one hand, given $S$, the space of all possible solutions for a problem $X$, a ``search'' problem is when we need to find one solution $s \in S$ for which a boolean predicate is evaluated as true, i.e., $f(s)=true$.
For example, $X$ could be the SE problem of automated bug fixing~\cite{arcuri2008automation,huang2024evolving}, where $S$ is the space of all possible patches, and $f(s)$ verifies if, for the given patch $s$, all current test cases pass (and so the bug can be considered as fixed).

On the other hand, in an ``optimization'' problem, the function $f(s)$ gives a numerical score. 
Depending on the problem, the objective would be to either \emph{minimize} or \emph{maximize} such a score. 
For example, in test data generation~\cite{mcminn2004search,ali2009systematic}, where $S$ is the space of all possible test cases for the given software under test, then $f(s)$ could be a metric like code coverage, where we try to find the test case $s_m \in S$ with maximum value for $f$.

For any non-trivial problem, the space $S$ of all possible solutions is extremely large (\sbseweakness). 
Given $n$, the length in bits of the representation of a solution, there are $2^n$ possible combinations.
This exponential number of combinations cannot be brute forced, i.e., enumerated fully. 
For example, consider something as simple as the testing of the function \texttt{classify(a,b,c)} to classify triangles from the length of their edges (a common example in the software testing literature~\cite{myers2004art,mcminn2004search,arcuri2008theoretical}).
There are four possible outcomes: scalene, isosceles, equilateral, and invalid. 
Given the three inputs as 32-bit integers, there are $|S|=2^{96}$ possible inputs for this function. 
That is more than 79 octillion possible combinations.

Given a reasonable amount of time, only an extremely tiny subset of $S$ can be evaluated using the function $f$. 
Brute force approaches in which each solution in $S$ is enumerated and evaluated with $f$ are infeasible in all but trivial problems. 
A \emph{search algorithm} will evaluate some solutions in $S$, compute their score $f(s)$, and then decide which other solutions in $S$ to evaluate. 
Such a decision is made with the so-called \emph{search operators} (e.g., do small random modifications to an already evaluated solution with a good $f$ score). 
This process is then repeated in a loop until either $k$ solutions have been evaluated, or a certain amount of time (e.g., 1 hour) has passed (i.e., based on the chosen \emph{stopping criterion}).
The longer a search algorithm is left exploring the search space $S$ by evaluating solutions in it using $f$, the better results can be expected.

There are many different search algorithms, including, for example, Hill Climbing~\cite{selman2006hill}, Genetic Algorithms~\cite{holland1992genetic}, and Ant-Colony Optimization Algorithms~\cite{dorigo1999ant}.
However, no algorithm is best on all possible problems~\cite{wolpert2002no}.
Mathematically, it is not possible. 
The more domain knowledge specific to a problem $X$ can be exploited by a search algorithm, the better results can be expected. 
This is one of the main drivers for designing ad-hoc, specific search algorithms tailored for the specific properties of the addressed problem $X$. 
An example is the Many Independent Optimization Algorithm for test suite generation~\cite{arcuri2018test}.

There are many problems in science and engineering that can be modeled as search or optimization problems~\cite{nocedal2006numerical,kramer2017genetic}. 
Traditional problems include, for example, the Traveling Salesperson Problem~\cite{rosenkrantz1974approximate}
and the Knapsack Problem~\cite{salkin1975knapsack}.
But they appear in all domains, from bioengineering to economics.
SE is just one of many kinds of fields that has tasks that can be modeled as search/optimization problems.
There are decades of success stories about the application of search/optimization algorithms~\cite{nocedal2006numerical,kramer2017genetic}.
Their success in addressing SE problems is, therefore, not unexpected~\cite{harman2012search}.

As for any technique, there are places and contexts in which search algorithms do not work well.
One of the major challenges in applying search to SE problems is the definition and design of the score function $f$ (\sbseweakness). 
This is problem-dependent and requires expertise to create one.
For example, assume the problem $X$ of generating test cases. 
On the one hand, you can try to optimize the generation of test cases to maximize metrics such as code coverage. 
This can be implemented in an $f$ function relatively easily (most programming languages already have existing libraries to measure code coverage). 
But what about \emph{code readability}? 
I.e., what about the generation of test cases that are easy to understand for humans? 
How to mathematically quantify in a function $f$ if a piece of code $s$ is easy to read and understand? 
If you cannot design and implement a score function $f$ for a given problem $X$, then you cannot use SBSE to address such a problem. 
For several tasks in SE, where, for example, natural language is involved (e.g., comments in Pull Requests, or text descriptions of product requirements), it might be challenging, if possible at all, to define a score function $f$. 

Another major issue is that, to be a viable option, the score function $f$ must not be too slow and resource-intensive to compute (\sbseweakness). 
A search algorithm, to provide good enough results, might need to evaluate tens or hundreds of thousands of solutions $s$. 
This, of course, depends on how long the search process is left running (i.e., depending on the chosen stopping criterion). 
For example, in the literature, experiments on unit test generation are run for something like 2 minutes~\cite{fraser2014large}, whereas experiments on system testing are run for 1 to 24 hours~\cite{golmohammadi2023testing}. 
If the score function $s$ only needs a few milliseconds to compute, then in these contexts the use of SBSE is feasible (as many success stories have shown~\cite{harman2012search}). 
However, what about the cases in which $f$ might take hours to compute? 
Even if $f$ takes ``only'' 1 minute to compute, at most only 1440 solutions could be evaluated in a single day. 
Furthermore, what about all other cases in which a response (i.e., finding the best $s$) must be provided quickly, e.g., in the order of a few (milli)seconds? 
In all those cases, likely SBSE techniques would not be particularly beneficial.

When talking about the execution time of $f$, there are two important aspects to consider: \emph{parallelization} and \emph{real-world constraints}. 
Advanced search algorithms are able to compute the score function $f$ in parallel. 
Instead of evaluating one solution $s$ at a time, a search algorithm could evaluate $n$ solutions in parallel, where $n$ could be based on the number of CPUs available on the machine on which the algorithm is running. 
Of course, this will impact how search operators are defined, but there are several different techniques to handle it (e.g., ``island models''~\cite{whitley1999island}). 
Based on the hardware available, parallelization can provide a linear speed-up to the search process. 
The possibility to automatically adapt to the available hardware resources is a major feature of search algorithms (\sbsestrength).

But, how much hardware is available?
This depends on the real-world scenarios in which these techniques are applied. 
In the case of SE problems, a common example is developers and testers using their work laptops for SBSE tasks. 
As of 2025, a typical laptop can have several CPU cores (e.g., not uncommon to have between 4 and 8 CPU cores).
For example, this very text is written on a Mac with an M4 chip having a 16-core CPU.
Considering the cost of hardware (e.g., laptops and servers) and electricity, compared to the salary cost of professional software engineers, the use of extra hardware dedicated to SBSE tasks can be warranted. This is the case if the use of SBSE brings a positive return on investment by saving (the expensive) time of software engineers (\sbsestrength).

Without disrupting existing software development processes, extra hardware that could be used for SBSE tasks might already be available.
For example, considering a common practice like the use of Continuous Integration (CI) servers~\cite{shahin2017continuous}, extending such servers to run SBSE tasks is possible (e.g., as done in~\cite{campos2014continuous,arcuri2025widening}). 
Different SE tasks, like test generation, bug fixing, and so on, could automatically be run in CI after developers make code changes (\sbsestrength).

In summary, SBSE can excel in all problems where a score function $f$ can be defined, and its computational cost is not too high. 
The longer a search process is left running, the better results can be obtained. 
SBSE techniques can be successfully run on developer laptops, as well as being able to exploit high parallelism when run on dedicated servers. 
Their adaptability to available hardware is one of their major strengths.

\subsection{Benefits and Downsides of FMs}
FMs represent a groundbreaking advancement in AI due to their ability to process multimodal data~\cite{bommasani2021opportunities}. 
These models encompass diverse architectures, such as LLMs for natural language processing and VLMs or MMs for handling image, audio, and video data~\cite{LMMsDawn2023}. 
A key strength of FMs is that they are trained using enormous and heterogeneous datasets, due to which they possess extensive cross-domain knowledge and the ability to identify complex patterns across different types of inputs (\fmstrength). 
FMs also feature task versatility due to their transformer-based architectures and self-supervised learning techniques. 
These capabilities enable them to understand context and provide outputs with reasoning that closely resembles human thinking, making them highly effective for complex decision-making and problem-solving tasks (\fmstrength). 

One of the most notable benefits of FMs is their adaptability and generalization in various domains (\fmstrength). 
Unlike traditional models that require extensive retraining for new tasks, FMs can adapt and generalize to domain-specific tasks with minimal fine-tuning. 
This adaptability has driven their adoption in numerous fields, including SE, autonomous systems, and robotics~\cite{hassan2024rethinking,firoozi2025foundation,gao2025foundation}. 
A recent advancement is vision–language–action (VLA) models~\cite{sapkota2025vla}, which extend the multimodal capabilities of FMs by not only interpreting complex inputs but also generating executable actions (\fmstrength). 
This makes VLAs particularly valuable in dynamic and interactive environments, such as autonomous vehicles or robotics, where they enable advanced control and embodied intelligence~\cite{sapkota2025vla}.

Despite their significant benefits, FMs also come with certain downsides. 
A primary concern is that these models often struggle in handling dynamic systems (\fmweakness), as they lack the real-time adaptability needed to respond to evolving requirements or changes in feedback loops~\cite{firoozi2025foundation}. 
For example, in applications such as autonomous systems, FMs lack the ability to dynamically incorporate new information.
Furthermore, their inherent limitations, such as hallucination, non-determinism, uncertainty, and output variability, raise serious reliability concerns (\fmweakness), especially in safety-critical domains such as robotics and healthcare, where factually incorrect outputs may cause safety risks~\cite{chakraborty2025hallucination,kawaharazuka2024real}.

Another significant challenge lies in the resource-intensive nature of FMs' fine-tuning and retraining for specific applications (\fmweakness). 
Customizing these models requires substantial computational resources, access to high-quality domain-specific data, and expertise, which may not be feasible for all applications~\cite{myers2024foundation}.
Moreover, the rapid evolution of FMs requires frequent updates and integrations for applications built on them, which pose challenges to long-term stability and compatibility (\fmweakness)~\cite{ran2025foundation}. 
In industrial settings, where trust and security are key concerns, FMs can introduce risks related to sensitive data and schema-related vulnerabilities (\fmweakness), which can limit their adoption~\cite{ran2025foundation}. 
Finally, ethical considerations of FMs (\fmweakness), such as bias, fairness, and the potential for misuse, further complicate their deployment in socially sensitive applications like finance~\cite{myers2024foundation}. 
These challenges emphasize the need for ongoing research to address the limitations of FMs and ensure their safe and responsible deployment.

\section{FMs for SBSE}\label{sec:RootFMs4SBSE}

\begin{figure}[t]
\centerline{\includegraphics[width=1\linewidth]{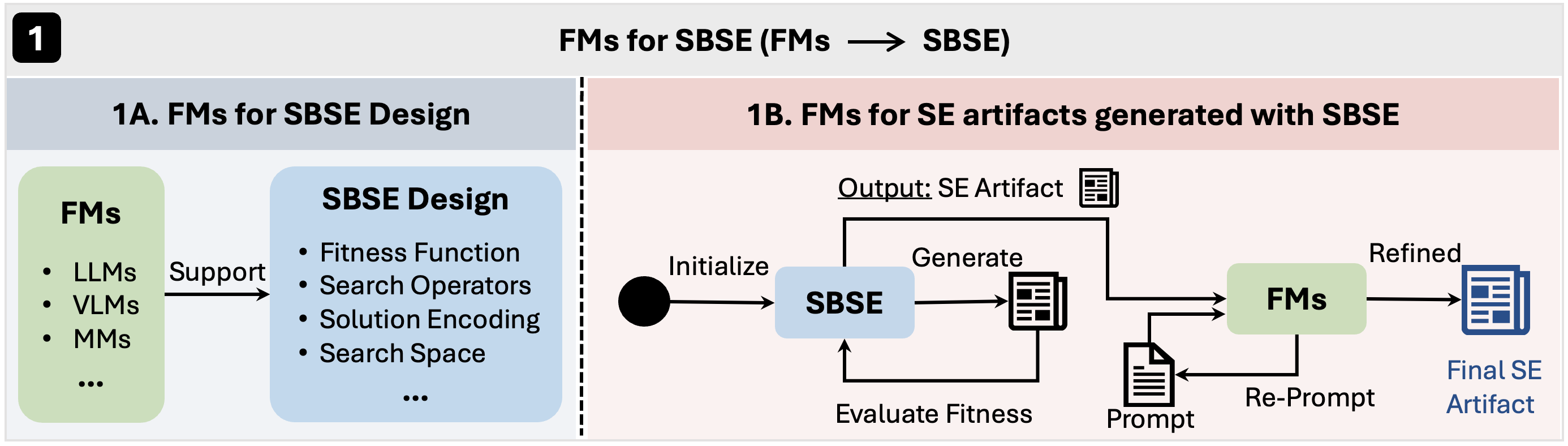}}
\caption{Key aspects of employing FMs to enhance SBSE. The abbreviations used are: FMs (Foundation Models), LLMs (Large Language Models), VLMs (Vision-Language Models), MMs (Multimodal Models), SBSE (Search-Based Software Engineering), and SE (Software Engineering).}
\label{fig:FM4SBSE}
\end{figure}

In this section, we discuss the role of FMs in enhancing SBSE, considering two sub-aspects: (1A) FMs to enhance SBSE design and (1B) FMs to enhance SE artifacts generated by SBSE, as illustrated in \Cref{fig:FM4SBSE}.
\begin{table}[!t]
\caption{\revision{Key challenges -- FMs for SBSE}}
\label{tab:keychallengesFM4SBSE}
\begin{tabular}{|p{0.99\textwidth}|}
\hline
\rowcolor[HTML]{E2EDF9}
\textbf{FMs for SBSE Design} \\ \hline
\ding{70} Automating the design and implementation of fitness functions for SBSE problems.\\
\ding{70} Designing search operators for complex domain problems involving multimodal data.\\
\ding{70} Automating solution encoding for the varying nature of problems.\\ 
\ding{70} Defining appropriate search spaces for SBSE to balance efficiency and solution quality.\\
\ding{70} Generating partial or complete SBSE implementations to enhance automation and efficiency.\\
\ding{70} Automating the testing, debugging, and repair of SBSE implementations.\\
\hline

\noalign{\vskip 0pt plus 0fil \break}\hline %for page break
\rowcolor[HTML]{E2EDF9}
\textbf{FMs for SE Artifacts Generated with SBSE} \\ \hline
\ding{70} Interpreting and transforming SBSE-generated artifacts into textual or graphical formats. \\
\ding{70} Static analysis of SBSE-generated artifacts for bug detection, test effectiveness, and program verification. \\
\ding{70} Supporting Pareto analysis and pruning for multi-objective SE problems, particularly for high-dimensional trade-offs and intuitive decision-making.\\ 
\ding{70} Validating and improving the quality of SBSE-generated artifacts, such as test scenarios, models, and code.\\ 
\ding{70} Context-aware input generation, flaky test detection, and repair in delta debugging. \\ 
\hline
\end{tabular}
\end{table}
In sub-aspect 1A, we analyze how FMs can automate aspects of SBSE design, such as solution encoding and fitness function design, which typically require domain experts' involvement. 
Conversely, in sub-aspect 1B, we explore how FMs can be applied to enhance SE artifacts generated by SBSE. 
For example, in search-based test data generation, while SBSE generates test data, creating platform- or language-specific executable tests involves embedding the test data with the test sequence and oracle, along with scripting for a given programming language and test execution platform. 
In such cases, FMs can complement search-based techniques by automating the generation of executable test scripts. 
The key distinction between 1A and 1B lies in their focus: 1A addresses the initial design phase of SBSE, while 1B emphasizes improving the SE artifacts generated by SBSE. 
\revision{For both aspects, a summary of the key challenges discussed in this section is presented in~\Cref{tab:keychallengesFM4SBSE}.}

\subsection{FMs for SBSE Design}\label{sec:FMs4SBSE}
In this section, we discuss the potential of FMs to advance SBSE design, starting with an overview of the current state of research, followed by a discussion of the key challenges and opportunities for further research.

\subsubsection{Current Landscape}

In the following, we present the current state of research focusing on SBSE applications throughout the SE lifecycle, the role of FMs in evolutionary computation, and FMs for the generation and optimization of search algorithms.

\paragraph{SBSE Applications across the SE Lifecycle.}
The typical SBSE process comprises several key steps, including problem formulation, solution encoding, search operator selection, definition of the search space, design of the fitness function, selection of the search algorithm, and specification of the stopping criteria~\cite{harman2001search}. 
Each of these steps requires careful consideration when applied to SE problems. 
Over the years, a substantial body of research has focused on the design of SBSE solutions for various phases of the SE lifecycle, from requirements engineering to testing and deployment~\cite{colanzi2020symposium,arcuri2013parameter}. 
For example, in requirements engineering, several efforts have been devoted to developing appropriate solution encoding schemes, designing specialized search operators, and crafting fitness functions to optimize requirements engineering problems such as requirement selection, prioritization, fairness analysis, and traceability~\cite{harman2012search}. 
For software design, the available literature has focused on various aspects such as defining search operators and fitness functions for model and architecture optimization~\cite{burdusel2021automatic,casamayor2023studying}. 
Another area that has gained significant attention is search-based software testing~\cite{mcminn2011search}. In this area, research has focused on devising novel mutation operators for cost-effective multi-objective test case selection~\cite{ugarte2025enhancing}, automatically inferring fitness functions for test case generation~\cite{formica2023search}, dynamically selecting fitness functions for unit test generation with high fault detection effectiveness~\cite{almulla2022learning}, reducing search space and reusing previous solutions for test data generation~\cite{sartaj2019search,sartaj2025search}, and improving quality indicators for multi-objective optimizations such as test selection~\cite{schutze2012using,panichella2014improving}. 
Similarly, in software debugging, several studies have focused on various aspects, including the selection of mutation operators and the refinement of fitness functions to enhance the effectiveness of automated program repair~\cite{guizzo2021refining,hanna2025reinforcement,huang2024evolving,repairSPLsJSS2024}.

\paragraph{FMs in Evolutionary Computation.}
Recently, LLMs have attracted significant interest in evolutionary computation due to their ability to enhance various aspects~\cite{wu2024evolutionary}. 
An important step in this regard is to use LLMs as search operators for single and multi-objective optimization, such as language model crossover (LMX) for single-objective~\cite{meyerson2024language} and LLM-assisted offspring generation for multi-objective optimization~\cite{wang2024large}. 
In the same line of work, \citet{freire2024large} investigated the use of LLMs to generate customized suggestions for objective functions aimed at optimizing product line architecture designs. 

\paragraph{Generation and Improvement of Search Algorithms.}
Another line of work is evolutionary algorithm generation and improvement using LLMs. 
Notable works include using LLMs to generate swarm intelligence algorithms~\cite{pluhacek2023leveraging}, designing heuristic algorithms~\cite{liu2024evolution}, and evolving search algorithms~\cite{ye2024reevo}. 
In addition, some studies also focused on using LLMs to assist novice users in interpreting and comprehending evolutionary algorithm-generated solutions~\cite{maddigan2024explaininggp,singh2024enhancing}. 
While LLM-based evolutionary computation has recently gained attention in the literature, its effectiveness within the SBSE context remains understudied. 
Given LLMs' potential to enhance evolutionary computation, we anticipate that other FMs, such as VLMs and MMs, could further complement SBSE in solving complex domain problems like autonomous systems~\cite{song2025generativeaitestingautonomous,gao2025foundation}, robotics, and quantum computing~\cite{qseRoadmapTOSEM2025}; for example, in the autonomous driving domain, LLMs have been combined with search algorithms to generate tests: LLMs first derive logical scenarios from reports of real accidents, and then multi-objective search for concrete scenarios in the search space defined by these logical scenarios~\cite{tang2024legend}.

\paragraph{Summary.}

An analysis of the current landscape indicates that key SE tasks addressed include \textit{software model optimization}~\cite{burdusel2021automatic}, \textit{fault detection in software models}~\cite{casamayor2023studying}, \textit{test case generation}~\cite{formica2023search,almulla2022learning}, \textit{test selection}~\cite{ugarte2025enhancing,schutze2012using,panichella2014improving}, \textit{test data generation}~\cite{sartaj2019search,sartaj2025search}, \textit{program repair}~\cite{guizzo2021refining,hanna2025reinforcement,huang2024evolving,repairSPLsJSS2024}, and \textit{product line optimization}~\cite{freire2024large}. 
Traditional SE tasks, such as software model optimization~\cite{burdusel2021automatic} and test selection~\cite{ugarte2025enhancing,schutze2012using,panichella2014improving}, primarily focus on key aspects of search problem formulation, like fitness function design, mutation operator selection, and search space reduction. 
In contrast, recent research is increasingly utilizing LLMs for advanced tasks, including algorithm generation~\cite{pluhacek2023leveraging}, heuristic design~\cite{liu2024evolution}, search operators~\cite{meyerson2024language}, and solution interpretation~\cite{maddigan2024explaininggp, singh2024enhancing}, thereby highlighting their growing role in SBSE.

\subsubsection{Roadmap}\label{ssec:roadmapfm4sbse}
Although the recent literature predominantly focuses on using LLMs (rather than the broader scope of FMs) for various aspects of SBSE design, numerous challenges persist, which present valuable opportunities for further research. 
In the following, we explore these open challenges and research opportunities in detail. 

\paragraph{Fitness Functions for SBSE Problems.}
Typically, a software engineer manually identifies and defines the fitness functions for SBSE problems based on their understanding of the domain (\fullref{SB-W2}). FMs, leveraging their strengths \fullref{FM-S1} and \fullref{FM-S2}, have the potential to assist in automating the definition and subsequent implementation of fitness functions for a given SBSE problem. For example, in test case prioritization, FMs can help provide alternative fitness function options to guide search algorithms in prioritizing test cases, e.g., based on historical data on what type of tests are more likely to detect faults. We foresee the following options to be investigated with FMs for designing fitness functions for SBSE problems:
\begin{inparaenum}[(1)]
\item FMs can be adapted as chatbots for SBSE users, who can build fitness functions by interacting with the chatbots and then implement the fitness functions themselves;
\item SBSE users can provide a set of requirements and ask the FM to implement the fitness function.
\end{inparaenum}
In addition, VLMs, SAMs, and MMs can be used to derive fitness function suggestions from images, audio, and video data, for example, when generating test scenarios for autonomous driving systems (ADS).

\paragraph{Solution Encoding for SBSE Problems.}
A fundamental step in SBSE is solution encoding for a particular SE problem, to identify the relevant search variables of the problem. 
This varies from one problem nature to another and requires problem-domain knowledge. 
FMs with diverse capabilities (\fullref{FM-S1} and \fullref{FM-S2}) can support solution encoding for different SE lifecycle phases. 
For instance, to verify software models with SBSE during the design phase, MMs with both textual and visual capabilities can facilitate solution encoding by analyzing software requirements and design artifacts, often created using graphical modeling languages such as Unified Modeling Language (UML)~\cite{uml}.

\paragraph{FMs for Search Space Definition.}
Most of the algorithms used in SBSE also require the definition of the search space, i.e., the range of possible values of the search variables (\fullref{SB-W1}). Having a too big search space can hinder the effectiveness of the search; on the other hand, a too small search space could exclude good solutions. FMs can provide a solution in this context by utilizing their strength \fullref{FM-S2}. For example, for test case generation, they could identify the areas of the program's input space that should be searched to increase the likelihood of generating tests that cover the entire program. Moreover, they could also identify the minimum granularity of value change that is meaningful, i.e., that can bring an observable change in the program output.

\paragraph{FMs for SBSE Implementation Generation.}
One advanced application of FMs (LLMs specifically) is code generation. While different approaches~\cite{surveyCodeGenerationTOSEM2025} have been proposed for LLM-based code generation, no approach is specific to generating implementations of SBSE problems. We forecast that LLMs (with their strengths \fullref{FM-S1}, \fullref{FM-S2}, and \fullref{FM-S3}) could be adopted to generate either partial or complete implementations of SBSE problems. In a partial implementation, LLMs can generate code to solve an SBSE problem, which the user then completes manually. Alternatively, LLMs can create complete implementations of SBSE problems that can be further refined.
This capability extends across key SBSE components, including solution encoding, operator selection, fitness function design, and search algorithm selection, enabling more efficient and automated problem-solving.

\paragraph{FMs for Testing, Debugging, and Repair of SBSE Implementation.}
The implementation of SBSE problems is challenging to test due to the absence of test oracles. To this end, FMs can use their analytical capabilities (\fullref{FM-S2}) to automate the testing of SBSE problems, ensuring that their implementation is correct. Therefore, further research is needed to assess whether FMs can be used to generate tests as well as test oracles for a given SBSE problem. Once issues are identified in SBSE implementations through testing, it is important to debug and repair them. These tasks are typically performed manually. In this context, FMs can be used to automatically debug SBSE implementations and generate patches to fix identified issues. However, this requires further research investigation, which has not been a focus in the literature.

\subsection{FMs for SE Artifacts Generated with SBSE}\label{sec:FMs4SEGenSBSE}
In this section, we examine the role of FMs in complementing SBSE to enhance SE artifacts. 
We begin with an overview of the current research landscape, followed by a discussion of open challenges and potential opportunities for future advancements.

\subsubsection{Current Landscape}

In the following, we explore literature focusing on FMs' role in improving SBSE-generated SE artifacts and FMs for Search-based REST API Testing.

\paragraph{Enhancing SBSE-generated SE Artifacts.}
In recent years, there has been growing research interest in various SE areas in using LLMs for SBSE. 
Software program refactoring represents another emerging direction where FMs are applied in SBSE. 
In this regard, \citet{choi2024iterative} utilized LLMs to guide a search-based approach, enhancing refactoring efficiency while reducing complexity. 
Another key area of interest is software testing, where LLMs are being explored for search-based software testing (SBST). 
In this case, \citet{turhan2024evollve} introduced evoLLve'M, an approach that utilizes LLMs to enhance JUnit test assertions generated by EvoSuite's evolutionary process~\cite{FraserTSE2013,FraserFSE11}. 
Similar to this work, \citet{biagiola2024improving} employed LLMs to improve the readability of unit tests generated by EvoSuite through search techniques.

\paragraph{Search-based REST API Testing.}
Search-based techniques have long formed the foundation of REST API testing, gaining significant research attention and leading to the development of robust tools such as EvoMaster~\cite{arcuri2019restful} and RESTler~\cite{atlidakis2019restler}. 
Some of these tools have even been applied in industrial contexts, for example, EvoMaster to test healthcare and automotive applications~\cite{sartaj2023testing,poth2025technology}. 
However, the industrial adoption of these search-based tools was often hindered by their limited capacity for text generation and comprehension~\cite{arcuri2025introducing}. 
Recently, LLMs, with their advanced natural language processing capabilities, have been increasingly customized and used for REST API testing, giving rise to a new generation of tools such as RESTGPT~\cite{kim2024leveraging} and LlamaRestTest~\cite{kim2025llamaresttest}. 
Similarly, in the context of evolving applications, search-based regression and differential testing of REST APIs have been initially explored in different domains~\cite{godefroid2020differential,isaku2023cost,sartaj2025restapi}. 
LLMs are now being used to perform differential testing on tests originally generated by search-based tools like EvoMaster~\cite{isaku2025llms}.
This shift represents a major advancement in REST API testing, demonstrating the role of LLMs in enhancing and complementing search-based approaches for modern, dynamic, and evolving applications. 

\paragraph{Summary.}
In summary, the review of the literature indicates that
\begin{inparaenum}[(i)]
\item LLMs, among various FMs, are increasingly used for SE tasks such as \textit{code refactoring}~\cite{choi2024iterative}, \textit{unit test improvement}~\cite{turhan2024evollve}, \textit{unit test readability}~\cite{biagiola2024improving}, and various types of software testing activities including \textit{REST API testing}~\cite{arcuri2019restful,atlidakis2019restler}, \textit{differential testing}~\cite{godefroid2020differential,isaku2025llms}, and \textit{regression testing}~\cite{isaku2023cost,sartaj2025restapi}, 
\item customized FMs like RESTGPT~\cite{kim2024leveraging} and LlamaRestTest~\cite{kim2025llamaresttest} are emerging for specialized SE tasks, particularly \textit{REST API testing}, and 
\item numerous SE phases, including requirements engineering, software design, and debugging, remain largely unexplored.
\end{inparaenum}

\subsubsection{Roadmap} \label{ssec:roadmapfm4seartifacts}
Despite SBSE's long-standing role in solving SE problems, several challenges persist, which we anticipate can be addressed through FMs in the future. Below, we highlight key opportunities and research directions.

\paragraph{FMs for SE Artifact Interpretation.} 
An SE problem in SBSE (e.g., requirements prioritization) is formulated using different solution encoding schemes, such as Binary or Tree-based encoding, depending on the problem's nature. 
The final output is generated as an encoded solution, often requiring manual effort for interpretation and transformation into a specific SE artifact.
This effort varies based on the SE lifecycle phase. 
For instance, in the case of requirements prioritization for the next release problem (NRP), the SBSE-generated prioritization sequence needs to be interpreted in textual requirements and utilized for NRP.
Similarly, software models generated with SBSE need to be converted to graphical models. 
In this context, FMs, leveraging their strengths \fullref{FM-S1} and \fullref{FM-S2}, can play a significant role in two directions.
First, LLMs with text comprehension and analytical capabilities can be used for SBSE-generated SE artifacts that are required to be interpreted and represented in a textual format. 
Second, VLMs and MMs can be employed to analyze and transform SBSE-generated SE artifacts into graphical representations for easier interpretation.

\paragraph{FMs for SE Artifact Refinement.} 
For some SE problems, SBSE-generated solutions do not need to be transformed; still, they require manual verification by domain experts. 
This is particularly evident in scenarios such as test data generation for primitive types and program improvement with genetic programming. 
In the case of test data generation, test effectiveness using metrics like coverage needs to be analyzed after running tests on the system under test. 
Similarly, programs generated with genetic programming either need to be compiled and tested, or they need to be manually examined using code inspection and walkthrough techniques. 
Recent studies have explored the use of LLMs for static analysis, particularly to detect bugs and security vulnerabilities within code~\cite{li2024enhancing,li2025irisllmassisted}, as well as to generate static analyzers capable of identifying bug patterns~\cite{yang2025knighter}. 
Beyond these applications, LLMs (supported by \fullref{FM-S2}) have significant potential in enhancing artifacts generated with SBSE. 
For the test data generated using SBSE, LLMs can conduct a static analysis utilizing the source code and test data to provide insights on test effectiveness, such as untested paths or incomplete coverage scenarios. 
In developing LLM-based static analysis techniques, we foresee the potential for devising novel test effectiveness metrics that complement these techniques. 
Furthermore, LLM-based static analysis (e.g., type checking and data flow analysis) can be further extended to verify programs generated with SBSE statically, with the potential to significantly reduce false positives through contextual reasoning, as demonstrated in recent studies~\cite{li2024enhancing,li2025irisllmassisted,yang2025knighter}.

\paragraph{FMs for Pareto Analysis and Pruning.}
In the case of multi- and many-objective SE problems, SBSE generates non-dominated or Pareto-optimal solutions in the objective space. 
A common practice is to analyze the resulting solutions and select the best one that meets a particular set of requirements. 
This process needs to be repeated whenever requirements change. 
In this regard, FMs with capabilities \fullref{FM-S1} and \fullref{FM-S2} can play a key role.
LLMs can support domain experts in understanding the trade-offs between different objectives and selecting the best solution for a given scenario. 
Moreover, for non-expert users, VLMs or MMs can construct a visual representation of the Pareto front, illustrating the set of all non-dominated solutions and offering suggestions to assist in selecting the most suitable option. 
In the case of pruning Pareto optimal solutions, which involves simplifying the solution set for easier decision-making, several challenges highlight the potential role of FMs. 
One significant challenge is managing high-dimensional problems with more than four conflicting objectives~\cite{petchrompo2022review}.
Another challenge is maintaining solution quality when pruning methods are applied, particularly in complex problems in domains such as robotics~\cite{victorica2023stability}. 
A further challenge lies in identifying the most influential variables from the pruned set and presenting the results in a way that enables intuitive and intelligent decision-making for users~\cite{white2025distilling}. 

\paragraph{FMs for Validation of the Generated Artifacts.}
FMs can be used to validate the generated artifacts from various perspectives. For example, in the case of test scenarios generated for cyber-physical systems like ADS (e.g., using genetic algorithms)~\cite{avfuzzerISSRE20,MOSAT_FSE22,EpiTester,REMAP,reqsViolADSsASE21,avoidCollICST2020,GambiISSTA2019,multAvoidCollGECCO2020}, FMs (leveraging their capabilities \fullref{FM-S3} and \fullref{FM-S4}) can help determine whether the generated scenarios are realistic. To this end, some works already aim to assess the realism of generated test scenarios (e.g., \cite{wu2024reality}). However, these works remain preliminary and are very specific to test scenarios generated using a single testing technique; therefore, more general methods are required. FMs can also apply their strengths \fullref{FM-S1} and \fullref{FM-S2} to improve the quality of generated test scenarios from various perspectives. For instance, they can increase the diversity of scenarios and support the automated removal of redundant ones. To this end, they can help determine the similarity (both syntactic and semantic) between two scenarios, further reducing duplication. Similarly, FMs can also be used to validate other generated SE artifacts, such as models or code produced with genetic programming. These artifacts can be further enhanced by FMs from both functional and non-functional perspectives (e.g., removing redundant model elements or optimizing code for a specific target hardware).

Automated debugging, such as delta debugging, represents a promising opportunity to leverage FMs, particularly LLMs, by using their strength \fullref{FM-S2}. 
Although search-based delta debugging has been well-studied in the literature~\cite{zeller2002simplifying}, several challenges exist. 
One challenge is that automatically generated test inputs often reduce overall test coverage due to limited system information~\cite{valle2022towards}. 
Another persistent challenge is the identification of flaky tests, which is complicated by code dependencies and limitations of traditional methods like abstract syntax trees to detect code changes~\cite{song2024c2d2}. 
Recent research has begun to explore LLM-based approaches for flaky test detection and repair~\cite{more2025analysis,chen2024generic}, yet the potential of LLMs in the context of delta debugging remains understudied. 
This opens up an exciting research opportunity to develop techniques based on coding LLMs for context-aware input generation and improved flaky test detection, potentially transforming the automated debugging process.

\section{SBSE for FMs}\label{sec:RootSBSE4FMs}

\begin{figure}[t]
\centerline{\includegraphics[width=1\linewidth]{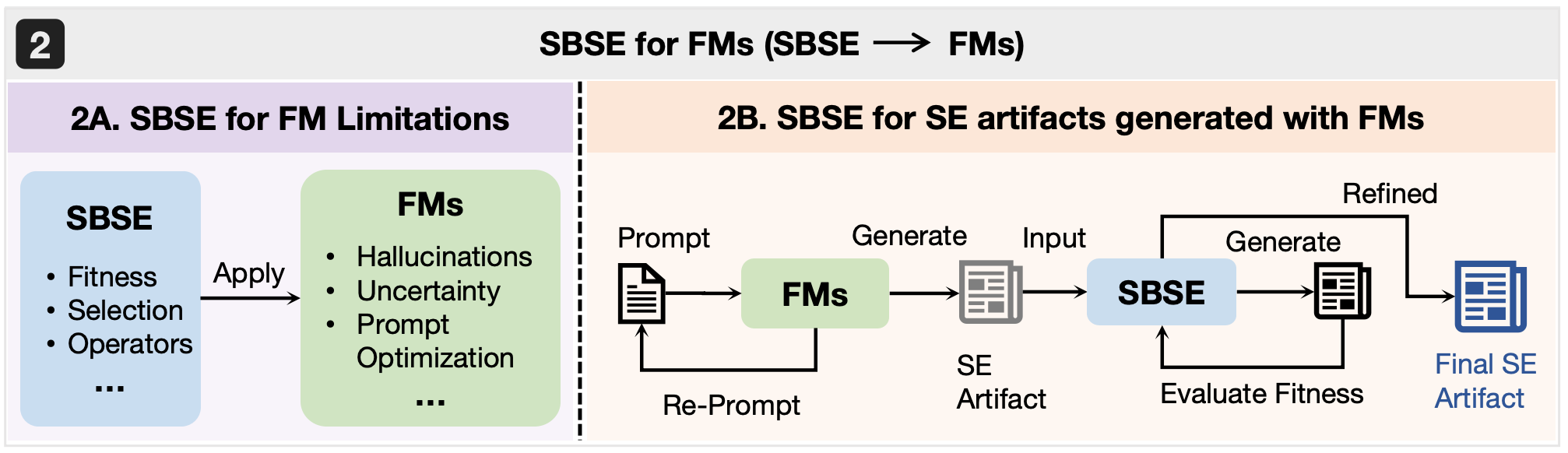}}
\caption{Key aspects of applying SBSE to enhance FMs. The abbreviations used are: FMs (Foundation Models), SBSE (Search-Based Software Engineering), and SE (Software Engineering).}
\label{fig:SBSE4FMs}
\end{figure}

In this section, we explore the role of well-studied SBSE techniques in enhancing FMs, considering two sub-aspects: (2A) using SBSE to address FM limitations and (2B) applying SBSE practices to enhance SE artifacts generated with FMs, as shown in~\Cref{fig:SBSE4FMs}.
For sub-aspect 2A, we analyze how SBSE can be applied to deal with the inherent limitations of FMs, such as hallucination and uncertainty. 
In contrast, for sub-aspect 2B, we explore how SBSE techniques can enhance SE artifacts generated by FMs, which are increasingly being customized for various SE tasks, such as coding LLMs for code generation. 
For example, in the context of coding LLMs, software genetic improvement can be applied to optimize the code generated by these models. 
The key difference between 2A and 2B lies in their focus: 2A addresses the inherent limitations of FMs, while 2B targets improving the outputs generated by FMs for SE tasks.
\revision{For both aspects, \Cref{tab:keychallengesSBSE4FM} summarizes the key challenges discussed in this section.  
}
\begin{table}[!t]
\caption{\revision{Key challenges -- SBSE for FMs}}
\label{tab:keychallengesSBSE4FM}
\begin{tabular}{|p{0.99\textwidth}|}
\hline
\rowcolor[HTML]{E2EDF9}
\textbf{SBSE for FM Limitations} \\ 
\hline
\ding{70} Managing non-determinism in FMs to ensure consistent and reliable outputs.\\
\ding{70} Effectively handling data-related uncertainties in FMs to improve model performance.\\
\ding{70} Detecting and mitigating hallucinations in FMs.\\ 
\ding{70} Optimizing the computationally expensive fine-tuning process of FMs, particularly for evolving SE requirements.\\ 
\ding{70} Balancing trade-offs in automated prompt optimization.\\ 
\hline

\rowcolor[HTML]{E2EDF9}
\textbf{SBSE for SE Artifacts Generated with FMs} \\ \hline
\ding{70} Refining FM-generated code to address compiler errors, resolve conflicts, and integrate into codebases.\\
\ding{70} Improving test quality, coverage, and diversity for FM-generated tests, while addressing flakiness and optimizing regression testing.\\
\ding{70} Enhancing FM-generated models (e.g., UML diagrams) for quality, correctness, and constraint validation.\\ 
\ding{70} Effectively optimizing FM-generated software configurations for functional and non-functional characteristics.\\
\ding{70} Debugging and repair of FM-generated artifacts.\\
\ding{70} Dealing with continuous evolution of FMs.\\
\hline
\end{tabular}
\end{table}

\subsection{SBSE for FM Limitations}\label{sec:SBSE4FM}
In this section, we explore the potential of SBSE in addressing the limitations of FMs. 
We begin with an overview of the current state of the art and then discuss the key challenges and opportunities ahead.

\subsubsection{Current Landscape}

In the following, we discuss research works focusing on SBSE for prompt engineering and the optimization of FMs applied throughout the SE lifecycle.

\paragraph{Prompt Engineering with SBSE.}
A growing area of research explores the application of SBSE techniques to optimize FMs-related activities, such as prompt engineering. 
LLMs, due to their advanced textual capabilities, have been widely applied in prompt engineering~\cite{wu2024evolutionary}. 
Recent efforts have targeted various aspects of prompt engineering, including optimizing textual prompts using search algorithms such as genetic algorithms and differential evolution to improve prompt quality~\cite{xu2022geneticpromptsearch,cui2025seestrategic}, applying evolutionary techniques for prompt tuning~\cite{guo2025evoprompt}, and employing multi-objective optimization methods to refine instructions within prompts~\cite{yang2023instoptima}. 
Another emerging direction is LLM security testing, where search-based techniques are used to generate adversarial prompts to identify vulnerabilities in LLMs~\cite{liu2024autodan,dang2025enhancingprompts}.

\paragraph{Optimizing FMs Applied across SE Lifecycle.}
In SE, a notable focus is on estimating software effort, an essential activity in software project management. 
For this purpose, search techniques have been applied to optimize few-shot learning methods for fine-tuning LLMs to improve their performance in estimating software effort~\cite{tawosi2023search}. 
Another focus area involves applying SBSE techniques to improve image generation quality, which has significant implications for tasks such as automated user interface design. 
In this direction, \citet{berger2023stableyolo} presented StableYolo, which applies a multi-objective search to optimize LLM prompts and parameters to generate high-quality and realistic images using tools such as YOLO and Stable Diffusion~\cite{berger2023stableyolo}. 
Building on this work, GreenStableYolo was introduced, which applies a multi-objective search for efficient LLM inference while maintaining image quality~\cite{gong2024greenstableyolo}. 
A recent study introduced search-based prompt refinement techniques to improve LLM-based code generation~\cite{taherkhani2025automated}.

\paragraph{Summary.}

The review of the literature reveals that the most commonly addressed SE tasks are \textit{software effort estimation}~\cite{tawosi2023search}, \textit{security testing}~\cite{liu2024autodan,dang2025enhancingprompts}, and \textit{code generation}~\cite{taherkhani2025automated}. 
A variety of search aspects, including genetic algorithms, differential evolution, and multi-objective optimization, have been applied to LLM-specific tasks such as prompt quality improvement~\cite{xu2022geneticpromptsearch,cui2025seestrategic}, prompt tuning~\cite{guo2025evoprompt}, and prompt refinement~\cite{yang2023instoptima}, as well as SE tasks like code generation, security testing, and software effort estimation. 
These early efforts highlight the potential of SBSE in addressing the limitations of FMs, extending beyond LLMs, and enabling FMs applications across diverse domains.

\subsubsection{Roadmap}\label{ssec:roadmapsbse4fms}
A key limitation noted in the literature is the lack of sufficiently large, high-quality datasets. 
Since FMs like LLMs are trained on enormous amounts of data, applying SBSE to optimize techniques such as prompting or few-shot learning necessitates access to similarly extensive and reliable data. 
Despite this challenge, an overview of existing works on applying SBSE to LLMs indicates that SBSE has significant potential for optimizing SE tasks utilizing FMs. 
Overall, this highlights several open problems that need to be addressed and significant opportunities for future research to expand the application of SBSE techniques to a broader range of FMs (e.g., VLMs and MMs).
Below, we present challenges and possible opportunities for further research.

\paragraph{Addressing FM Challenges.} 
A key challenge of FMs is non-determinism, where the same input can produce varying outputs, affecting consistency and reliability (\fullref{FM-W2}).
This non-determinism leads to several challenges in applying FMs to real-world use cases. 
For instance, in continuous integration and continuous delivery (CI/CD) and DevOps, variability in generated configurations may lead to inconsistencies in software releases~\cite{sartaj2025restapi}. 
Search algorithms, especially those designed to operate under noise and uncertainty~\cite{klikovits2023trust,dinu2025adaptive}, along with the SBSE strength \fullref{SB-S3}, offer promising means to manage non-determinism in FMs. 
One potential research direction is the development of methods to systematically explore input spaces to identify scenarios where model outputs exhibit high variability. 
This could involve building on traditional techniques such as noise-tolerant fitness functions, sampling strategies, and robust selection mechanisms~\cite{rakshit2017noisy}. 
Another promising direction for further research is the application of search techniques in identifying optimal configurations for software and hardware resources that minimize variability in model behavior.

\paragraph{Managing Uncertainty in FMs.}
Uncertainty is an inherent challenge of FMs (\fullref{FM-W2}).
Since FMs are trained on vast amounts of data, fine-tuning them also requires a significant amount of data, and the quality of such data (e.g., redundant features or imbalanced data) can be a potential source of uncertainty. 
For example, VLMs trained/fine-tuned on images containing poor contrast, low brightness, or blurriness can significantly affect model interference. 
Such uncertainties can significantly affect model interference, rendering these models ineffective for several domain-specific tasks. 
SBSE, leveraging strengths \fullref{SB-S1} and \fullref{SB-S2}, can be applied to select an optimal set of features, which will result in effectively managing data-related uncertainty and efficient training/fine-tuning of FMs.

\paragraph{Hallucinations in FMs.}
A key challenge in FMs is hallucination (\fullref{FM-W2}), where the model generates entirely fabricated or factually incorrect information~\cite{rawte2023surveyhalluc}, posing a significant concern in FM applications. 
As a result, detecting and mitigating hallucinations has become a key focus of ongoing research~\cite{chakraborty2025hallucination}. 
In this regard, we foresee significant potential for SBSE to effectively address the challenge of hallucinations in FMs. 
Specifically, SBSE can complement black-box methods for hallucination detection, as highlighted by the need for such methods~\cite{chakraborty2025hallucination}. 
Using a search-based approach with capability \fullref{SB-S1}, a search algorithm can iteratively generate a diverse population of prompts, collect corresponding responses from FMs, and employ a fitness function to continuously assess response variations, enabling more precise hallucination detection.

\paragraph{Fine-tuning FMs.}
FMs adapted for different SE phases are typically required to be fine-tuned using methods such as few-shot learning and transfer learning. 
With the continuous evolution of software following SE development models, FMs used in these phases are required to be fine-tuned repeatedly (\fullref{FM-W1}, \fullref{FM-W3}, and \fullref{FM-W4}). 
For instance, whenever software requirements change, the FMs for that particular software need to be fine-tuned to adapt to updated requirements. 
Given the size of FM parameters and modality, fine-tuning can be highly computationally expensive. 
In such cases, SBSE, applying strengths \fullref{SB-S1} and \fullref{SB-S2}, can be used to optimize the FMs' fine-tuning process, including optimizing various activities like selecting hyperparameters and model configurations. 
This could result in increasing the overall efficiency and streamlining FMs' adaptation in the SE development lifecycle.

\paragraph{Prompt Optimization.}
The performance of FMs is heavily influenced by the design of prompts, which are specified in different formats, including textual, visual, or structured formats like domain-specific languages~\cite{gu2023systematicsurvey}. 
However, manual prompt engineering is often labor-intensive and prone to errors (\fullref{FM-W2} and \fullref{FM-W5}), which has led to significant interest in automated prompt engineering~\cite{li2025surveyautoprompt}. 
Given the vast search space for identifying optimal prompts, prompt optimization has become an active research area~\cite{liu2025promptcontent}. 
In this regard, several techniques have been explored, including gradient-based methods~\cite{zhong2025gradpromptopt}, LLM-assisted optimization~\cite{manas2024improvingtext}, example-guided strategies~\cite{yan2025efficient}, and reinforcement learning~\cite{li2025surveyautoprompt}. 
Evolutionary techniques have also been applied, such as genetic algorithms for few-shot textual prompts in LLMs and evolutionary algorithms for one-shot prompts in VLMs~\cite{li2025surveyautoprompt,qu2025proapo}. 
Despite these advancements, a key challenge remains in balancing trade-offs among objectives such as high generation costs, longer execution times, and the risk of overfitting~\cite{qu2025proapo}. 
A promising research direction is applying SBSE capabilities \fullref{SB-S1} and \fullref{SB-S2} to cast prompt optimization objectives into multi- or many-objective search problems and analyzing search algorithms' effectiveness in exploring trade-offs. 
This will further open pathways to the development of novel search operators and heuristics tailored for automatic prompt optimization.

\subsection{SBSE for SE Artifacts Generated with FMs}\label{sec:SBSE4FMSE}
In this section, we analyze the role of SBSE in improving SE artifacts generated with FMs. 
We begin with an overview of the current research landscape, followed by a discussion of open challenges and potential opportunities for future advancements. 

\subsubsection{Current Landscape}

In the following, we explore recent research efforts on customizing and applying FMs to SE tasks and tackling the associated FM challenges.

\paragraph{Customizing and Applying FMs to SE Tasks.}
FMs are increasingly being integrated into various phases of the SE lifecycle, significantly transforming how SE activities are performed~\cite{li2025sefms,llmsForSEZhang2023arxiv}. 
This integration has led to the development of FMs tailored for specialized SE tasks, enabling more efficient and intelligent automation throughout the lifecycle. 
Prominent examples include LLMs tailored for various coding tasks, such as CodeX~\cite{chen2021evaluatingllms} designed for code generation and CodeGen~\cite{nijkamp2023codegen} for program synthesis.
In addition to code generation, LLMs have also been used in software testing to automatically generate test cases, significantly improving the efficiency of testing processes~\cite{fan2023large}. 
For debugging, LLMs have demonstrated usefulness for tasks such as fault localization, where they assist in identifying the root cause of software failures~\cite{wu2023llmfault,YangICSE24,LiuQRS24}. 
Furthermore, these models have been used to reproduce software bugs by interpreting bug reports and generating the corresponding scenarios using advanced prompting techniques~\cite{kang2024evaluating}.

\paragraph{FM Challenges in SE.}
As LLMs are increasingly used for various SE tasks, there are also several challenges. 
For the code and test cases generated using LLMs, evaluating their correctness is a key challenge~\cite{chen2021evaluatingllms,nijkamp2023codegen}. 
In addition, since FMs (including LLMs) are trained using open-source code and datasets, they often suffer from a lack of diversity, which is necessary for tasks such as fault localization and program synthesis~\cite{wu2023llmfault}. 
Another challenge is that FMs have no direct control over the execution environments of code and tests, thereby limiting their ability to support activities such as bug reproduction~\cite{kang2024evaluating}. 
To address these challenges, the SBSE community has begun to apply search optimization techniques to the SE artifacts generated with FMs.

\paragraph{Summary.}

An analysis of the current landscape indicates that core SE tasks being addressed include \textit{code generation}~\cite{chen2021evaluatingllms}, \textit{program synthesis}~\cite{nijkamp2023codegen}, \textit{test case generation}~\cite{fan2023large}, \textit{fault localization}~\cite{wu2023llmfault,YangICSE24,LiuQRS24}, and \textit{bug reproduction}~\cite{kang2024evaluating}. 
While the literature predominantly focuses on utilizing FMs (mainly LLMs) for various SE tasks and highlights challenges to be addressed, the application of search aspects in tackling these challenges has received limited attention. 
This indicates significant opportunities for SBSE to address these challenges and drive advancements in the area.

\subsubsection{Roadmap}\label{ssec:roadmapsbse4fmseartifacts}
Despite significant efforts to apply SBSE to enhance SE artifacts generated with FMs, several challenges remain, highlighting the need for further research opportunities. 

\paragraph{Search-based Code Improvement.}
Recent literature has shown significant interest in using general and fine-tuned LLMs for code generation.
However, LLM-generated code often leads to compiler errors (potentially due to weakness \fullref{FM-W2}), and testing such code remains an open challenge~\cite{fan2023large}. 
Furthermore, integrating LLM-generated code in the software code base requires significant manual effort in analyzing code, resolving conflicts, and testing the integrated code. 
The current practice to improve the code generated by LLMs is using prompt engineering~\cite{fan2023large}. 
Building on the extensive research and practical applications of search-based program improvement and code repair, applying these techniques to refine LLM-generated code opens significant research opportunities. 
This advancement will extend the applicability of traditional SBSE techniques from human-written code to optimizing LLM-generated code.

\paragraph{Search-based Testing of FMs-Generated Artifacts.}
Similar to code generation challenges, the test cases generated from LLMs suffer from several issues due to the limitation \fullref{FM-W2}, such as erroneous tests, low coverage, and test flakiness~\cite{fan2023large}. 
The existing approaches tackle these issues using different LLMs and prompting techniques. 
Given the well-studied SBST techniques, one potential research direction is to apply SBST (leveraging SBSE strength \fullref{SB-S1}) to improve tests generated with LLMs. 
Moreover, in case a large number of tests are generated with LLMs, it is infeasible to run all tests for regression testing in CI/CD or DevOps environments~\cite{sartaj2025restapi}. 
Therefore, another research direction is to apply SBST (utilizing \fullref{SB-S3}) for regression test selection, minimization, and prioritization techniques to LLM-generated tests. Moreover, during test optimization, search-based testing can improve test diversity by removing redundant test cases. Such research opportunities still exist and require further investigation in the coming years.

\paragraph{Search-based Optimization of Models Suggested by FM.}

Models generated with FMs, such as various UML diagrams, can potentially be further optimized using search algorithms while incorporating SBSE strength \fullref{SB-S1}. For instance, when FMs suggest multiple model variants, search-based techniques could be employed to select the best one based on defined fitness functions. Moreover, constraints defined on models may not be easily tested by FMs alone, and search-based testing techniques can help ensure their correctness. Although methods for such testing exist~\cite{jilani2022automated,boussaid2017survey}, their integration with FMs remains largely unexplored, and further studies are needed to systematically investigate how these combined approaches can enhance model quality and correctness.

\paragraph{Search-based Optimization of Configurations Suggested by FM.}
Recent studies have started exploring FMs, specifically LLMs, in identifying software configurations; however, there are several performance challenges~\cite{spieker2025prompting}. 
While search algorithms offer a promising approach to optimizing FM-generated software configurations (e.g., test, deployment, or software product line configurations), more work is needed to fully explore their potential in this context. Future research should investigate the effectiveness of different search techniques (using \fullref{SB-S1} and \fullref{SB-S2}), the impact of various fitness functions on functional and non-functional characteristics of software, and the integration of FM-generated initial configurations as an initial population for search-based techniques. Such studies could provide deeper insights into how these combined approaches can systematically improve software quality and performance by optimizing software configurations.

\paragraph{Search-based Debugging, Repair, and Evolution of FMs-Generated Artifacts.}
LLM-based fault localization and reproduction techniques generate fault reports, which are subsequently used for tasks such as debugging, program repair, and test evolution. 
Given that SBSE literature has well-established techniques for software maintenance and evolution, we argue that these techniques can be applied using SBSE strengths \fullref{SB-S1} and \fullref{SB-S2} to enhance the process.
For instance, search-based program repair techniques can be adapted to improve code as a next step after fault localization by LLMs. 
Moreover, once the code has been improved, another direction involves using SBSE (with \fullref{SB-S3}) to evolve tests based on the fault information.

\section{Integrating SBSE and FMs}\label{sec:SBSE&FMs}
\begin{figure}[t]
\centerline{\includegraphics[width=1\linewidth]{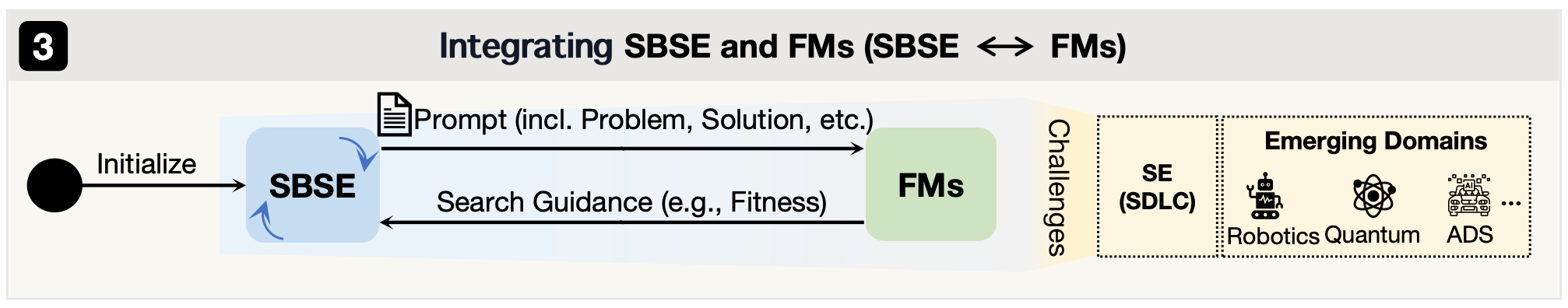}}
\caption{Integration between FMs and SBSE (two-way interactions). The abbreviations used are: FMs (Foundation Models), SBSE (Search-Based Software Engineering), SE (Software Engineering), SDLC (Software Development Lifecycle), and ADS (Autonomous Driving Systems).}
\label{fig:SBSE&FMs}
\end{figure}

In this section, we analyze the integration between SBSE and FMs, 
focusing on how SBSE and FMs can work together to leverage the strengths of both in addressing challenges across the SE development lifecycle and in emerging domains, as illustrated in~\Cref{fig:SBSE&FMs}.
Unlike \Cref{sec:RootFMs4SBSE} (FMs for SBSE) and \Cref{sec:RootSBSE4FMs} (SBSE for FMs), which focus on unidirectional benefits, this section discusses the two-way interaction between SBSE and FMs, exploiting the combined power of both to address complex problems. 
\revision{An overview of the key challenges discussed in this section is provided in~\Cref{tab:keychallengesSBSEandFM}.}
\begin{table}[!t]
\caption{\revision{Key challenges -- Integration of SBSE and FMs}}
\label{tab:keychallengesSBSEandFM}
\begin{tabular}{|p{0.99\textwidth}|}
\hline
\rowcolor[HTML]{E2EDF9}
\textbf{Integrating SBSE and FMs} \\ \hline
\ding{70} Developing FM-inspired search algorithms to enhance initialization, fitness calculation, and solution repair during the search process.\\
\ding{70} Using FMs to support SBSE in traditional SE problems, such as requirement prioritization, fault identification, and repair for SE tasks, including GUI testing.\\
\ding{70} Effectively integrating FMs with SBSE to tackle complex problems in emerging domains.\\ 
\hline
\end{tabular}
\end{table}

In the following subsections, we begin with an overview of the current state of the art and then discuss the key challenges and opportunities for further research.

\subsection{Current Landscape}

In the following, we discuss recent research literature on integrating SBSE and FM in various areas, including software development, genetic improvement, and search-based software testing.

\paragraph{Software Development.}
An increasing research trend is to explore the synergy between SBSE and LLMs, investigating how LLMs can enhance SBSE techniques and how SBSE can improve LLMs for SE tasks. 
One notable focus is software development, where \citet{gao2024search} introduced SBLLM, a framework for code optimization that leverages LLMs to apply evolutionary search techniques like crossover and mutation. 
Similarly, \citet{bouras2025llm} presented LLM-based crossover operators for genetic improvement of the source code. 
Moreover, \citet{wang2025large} proposed a masking mutation operator for genetic improvement that uses LLM code completion abilities for context-aware code changes during the mutation process.

\paragraph{Software Genetic Improvement.}
Another area is software genetic improvement (GI), where \citet{kang2023towards} presented the initial concept of generating mutants from LLMs to optimize the process of GI. 
Similar to this work, \citet{brownlee2025large} employed LLMs as mutation operators to increase diversity in the search process for GI. 
In the case of automated program repair, \citet{MurtazaGECCO2024} presented SemiAutoFL, a tool that combines genetic programming with LLMs to improve fault localization and bug fixing. 
For the application-specific use of search and LLMs, \citet{shojaee2024llmsr} presented LLM-SR, an approach that combines LLMs and evolutionary search to discover mathematical equations from the programs, demonstrating the potential to enhance the accuracy and efficiency with both LLMs and search techniques.

\paragraph{Search-based Software Testing.}
Another prominent area of interest is exploring the potential of LLMs with SBST. 
In this direction, \citet{dakhama2023searchgem5} introduced SearchGEM5, an approach that combines search-based fuzzing with LLMs to generate test cases for the Gem5 simulator. 
Building on this work, \citet{dakhama2025enhancing} introduced SearchSYS, an automated approach and tool designed to generate diverse test inputs using LLMs and code mutants through search-based fuzzing. 
Later, \citet{bruce2024search} presented an approach that utilizes LLMs to guide search-based fuzzing toward finding faults in ARM simulator software. 
\citet{lemieux2023codamosa} presented a technique called CodaMosa, which integrates a multi-objective search algorithm with a coding LLM to generate tests for Python code capable of achieving high coverage. 
Furthermore, \citet{sapozhnikov2024testspark} presented TestSpark, a tool that utilizes the search-based capabilities of EvoSuite and LLMs' textual capabilities to generate JUnit tests. 
Subsequently, \citet{abdullin2025testwars} conducted a comparative study of search-based tools, symbolic execution, and TestSpark configured with various LLMs. 
Their results demonstrated that while LLM-based tools underperformed in the case of code coverage, they outperformed in achieving higher mutation scores.
For automated program repair, \citet{lijzenga2025leveraging} integrated SBSE with coding LLMs to generate and validate code patches.
In the ADS domain, \citet{tang2024legend} used LLMs to derive logical scenarios from accident reports and then applied multi-objective search to generate critical test scenarios; \citet{YouISSRE2025}, instead, proposed a fault injection approach to generate critical scenarios that adopts LLMs to find realistic and diverse faults.

\paragraph{Summary.}

The overview of current literature reveals that key SE tasks primarily addressed are \textit{code optimization}~\cite{gao2024search}, \textit{code improvement}~\cite{bouras2025llm,wang2025large,kang2023towards,brownlee2025large}, \textit{program repair}~\cite{MurtazaGECCO2024,lijzenga2025leveraging}, \textit{test generation}~\cite{dakhama2023searchgem5, dakhama2025enhancing,lemieux2023codamosa,sapozhnikov2024testspark,tang2024legend}, and \textit{fault localization}~\cite{bruce2024search}. 
It is also observed that LLMs are being customized to act as search operators, such as crossover and mutation, or are utilized during the search process for various SE tasks, including code optimization, program repair~\cite{MurtazaGECCO2024,lijzenga2025leveraging}, test generation~\cite{dakhama2023searchgem5, dakhama2025enhancing,lemieux2023codamosa,sapozhnikov2024testspark,tang2024legend}, and fault localization~\cite{bruce2024search}. 
While current efforts primarily focus on LLMs, exploring the integration of advanced FMs (such as VLA/VLMs) with SBSE presents promising opportunities for addressing complex SE challenges and tackling problems in emerging domains like robotics and ADS.

\subsection{Roadmap}\label{ssec:roadmapfm&sbse}
While the integration of SBSE and FMs remains largely unexplored, early research reveals a promising synergy. 
However, it also highlights a significant performance limitation due to the substantial computational resources required by the integrated use of SBSE and FMs. 
In the following, we present a research roadmap and opportunities, exploring the potential of SBSE and FMs to advance metaheuristic search in SBSE and solve challenges in SE and emerging domains.

\subsubsection{Advancing Metaheuristic Search in SBSE}\label{subsec:advsearchinsbse}
As highlighted in the latest survey~\cite{wu2024evolutionary} and our literature analysis, FMs, specifically LLMs, can support various search algorithms in multiple ways. 
This includes using FMs to support single and multi-objective optimizations, act as fitness functions, and perform crossover and mutation.

\paragraph{Advancing Search Algorithms.}
Given the advantages that FMs bring to the search process, one potential research direction is developing novel FM-inspired search algorithms. 
Furthermore, search-based algorithms, such as genetic algorithms, initiate the search process with a randomly generated population. 
During the search process, different selection strategies like tournament and rank-based selection are used to pick individuals from the current population to generate offspring for the next generation. 
For solving complex domain problems, intelligently selecting and reusing solutions individuals have demonstrated to be an efficient strategy~\cite{sartaj2019search,sartaj2025search}. 
In this context, a promising research direction is using FMs' strengths \fullref{FM-S1} and \fullref{FM-S4} to leverage domain knowledge for generating an optimized initial population, thereby enhancing the efficiency of the search process. 
For instance, for SBSE applied to emerging domains requiring mixed-mode capabilities, such as robotics and ADS, FMs like VLMs or MMs can support generating the initial population. 
This could significantly broaden SBSE's applicability to diverse, complex domains with cutting-edge challenges.

\paragraph{Advancing Fitness Functions.}
During the search, FMs, with their capabilities \fullref{FM-S2} and \fullref{FM-S3}, can be used to calculate fitness, depending on the SBSE problem at hand. 
For example, in ADS testing, FMs can be queried to assess whether a scenario is realistic~\cite{wu2024reality}. 
This information can then be used to define fitness functions that guide search algorithms for SBSE problems. Thus, more research is needed to investigate which SBSE problems can utilize FMs to calculate fitness during the search. 
Another significant challenge is that search algorithms often generate invalid solutions that require repairs.
To address this, FMs can repair solutions during the search, ensuring the search is properly guided. To this end, new research methods based on FMs are needed to prompt FMs to repair solutions when issues are encountered during the search.

\paragraph{Advancing Search Operators.}
In a recent study, \citet{wu2024evolutionary} categorized various LLM-based search operators, such as crossover and mutation, for single- and multi-objective algorithms. 
However, the application of such operators in SBSE remains undetermined. 
Therefore, one potential research direction is to explore LLM-based search operators in the SBSE context, utilizing capabilities like \fullref{FM-S2}. 
Another research direction is to determine search operators at runtime using FMs' vision and multimodal capabilities \fullref{FM-S4}, and apply SBSE to solve problems for dynamic systems. 
For example, when dealing with ADS data in image format, VLMs can be used to perform crossover and mutation for the generation of image-based test scenarios~\cite{gao2025foundation}. 
Similarly, for ADS data in audio or video formats, search operators based on SAMs and MMs can enhance test scenarios considering the mixed-mode nature of data.

\subsubsection{Solving SE Problems}\label{subsec:solvingseproblems}
SBSE has been applied to many software requirement engineering (RE) problems, including requirement selection and prioritization, requirement satisfaction, and fairness analysis in requirements. 
Software requirements are typically elicited and documented in a textual format. 
FMs, specifically LLMs with text generation, comprehension, and analysis capabilities (\fullref{FM-S1} and \fullref{FM-S2}), can complement search-based RE in many ways. 
For instance, in the case of requirement selection, LLMs can be provided with software requirements and then act as a fitness function during search optimization to solve the next release problem. 
Moreover, for requirements prioritization with multi-objective search, LLMs can be used to analyze conflicting objectives and select Pareto-dominant solutions.

In addition to RE, SBSE techniques can be applied to test FMs used to solve SE problems (e.g., code generation and test generation). Taking the example of FMs generating source code, the aim is to test whether FMs generate correct code. For this purpose, SBST can be applied (leveraging SBSE strengths \fullref{SB-S1} and \fullref{SB-S2}) to generate test cases that prompt FMs to generate source code. The generated source code can then be used to compute a fitness function, guiding the search to identify faults that cause FMs to generate the wrong code. To this end, with carefully designed fitness functions, search-based testing techniques can be used to identify faults in FMs for various SE tasks. Once faults are identified using SBSE approaches, they can also be employed to repair FMs. This can be achieved by defining a fitness function to generate patches (e.g., fine-tuning parameters) for FMs until the faults are resolved.

In software testing, graphical user interface (GUI) testing presents a compelling application for integrating SBSE (involving \fullref{SB-S1} and \fullref{SB-S2}) and FMs (utilizing \fullref{FM-S1} and \fullref{FM-S4}). 
Modern software applications are designed with advanced GUIs to facilitate user-friendly interactions. 
GUI testing is essential to ensure functionality, visual consistency, compatibility, and usability. 
The available tools for GUI testing, like GUITAR~\cite{nguyen2014guitar} and Selenium~\cite{bruns2009selenium}, require test scripts to execute tests. 
Therefore, combining SBST with MMs, which possess mixed-mode capabilities, such as textual and graphical elements, presents an intriguing direction for future research.

\section{Challenges of Empirical Evaluations}\label{sec:empiricalchallenges}

When proposing a novel technique to address an SE problem, typically experiments are carried out on a set of artifacts, with comparisons made with the current state-of-the-art. 
If better results are obtained, then a novel contribution to the current body of knowledge has been made. 
However, to ensure that a novel technique $X$ is indeed better than the current state-of-the-art $Y$, experiments need to be \emph{fair}.

Typically, in SBSE experiments, and in experiments with search algorithms in general, compared techniques are run with the same resources (e.g., 1 CPU core) for the same \emph{search budget}. 
This could be a timeout (e.g., 2 minutes or 1 hour), or the same number of evaluated solutions (recall SBSE limitation \fullref{SB-W1} in Section~\ref{subsec:sbse}).
In other words, algorithms and techniques/tools are compared by trying to use the same amount of computational resources. 
If you do not have fair comparisons, you cannot really claim that a novel technique is better than the current state-of-the-art.

When using FMs or combining FMs with SBSE techniques, the same requirements still hold: experiments must be fair. 
However, how to ensure that the same amount of resources is employed in the case of FMs?
If you use a commercial FM, typically, you will be making an API call over the internet by sending a prompt as a payload, which introduces risks such as \fullref{FM-W5} and \fullref{FM-W6}. 
This payload would then trigger computations on a remote server, usually running on high-end GPUs. 
Then the response would be sent back. 
Such round-trip communication toward a remote server takes a non-zero amount of time.
Furthermore, what are the specs of those high-end GPUs? 
How much electricity did they consume to answer the prompt? 
How do those values compare with the specs of a consumer-grade laptop where you can run SBSE techniques? 

If a remote, commercial FM requires $N$-times the electricity and hardware capabilities compared to an SBSE technique, a more \emph{fair} comparison would be to run the latter in parallel with $N$ CPU cores. 
If a laptop/server does not have enough CPUs, then, as commercial FMs are run on remote servers, you could rent cloud resources (e.g., from Amazon Web Services or Azure) to run the SBSE technique in parallel.
If you can pay money for a subscription to use a commercial FM (e.g., ChatGPT), then you can use the same amount of money to rent cloud resources to run an SBSE technique in parallel. 
Still, how to reliably determine a fair value for $N$?

The use of remote, commercial FMs is problematic in this context due to potential risks involving \fullref{FM-W1}, \fullref{FM-W4} and \fullref{FM-W6}.
An alternative would be to use an open-source FM that can be run locally (e.g., via \emph{ollama}\footnote{https://ollama.com/}), 
as it makes evaluating its computational costs more feasible. 
However, without the proper hardware (e.g., GPUs), results could be quite poor compared to commercial, remotely run FMs. 
The need for specialized, more expensive hardware (compared to a consumer-grade laptop) is not necessarily a showstopper. 
Such costs could be justified based on the improvements obtained by solving the SE problems at hand.
Still, how to make proper fair comparisons when different techniques might require different kinds of hardware, e.g., GPUs vs.~CPUs? 
There is no easy answer here, apart from possibly checking their power consumption, which is not necessarily trivial to infer reliably.

However, the use of a local FM solves another critical issue: \emph{replicability}. 
A remote, commercial FM could change at any time (\fullref{FM-W4}), like any other service provided on the internet. 
Given an experiment involving such a remote FM, it might not be possible to replicate such a study if the FM has changed meanwhile~\cite{sallou2024breaking}.
Lack of replicability is a major issue in scientific research~\cite{national2019reproducibility,vcrepinvsek2014replication}. 
A result that cannot be replicated has limited scientific value. 
On the other hand, if the FM is an open-source one that can be run locally, experiments can be attempted to replicate, as long as similar hardware is employed. 
This does not mean that commercial FMs should not be used in scientific research, especially if they are the ones that give the best results.
Still, for sound results, experiments should be run as well with a local FM, which can then be used as a point of reference in replicated studies and comparisons.

Note that the concept of ``replicability'' is different from the concept of ``non-determinism''. 
If you prompt an FM twice, you might get different results even if the prompt is exactly the same. 
But this is exactly the same issue as for SBSE techniques, which most of the time are based on randomized algorithms. 
In those cases, experiments need to be repeated several times (e.g., 10 or 30 times), and the appropriate statistical tests should be used to analyze the results~\cite{arcuri2014hitchhiker}. 

How to properly compare, in a fair manner, hybrid techniques involving both FMs and SBSE techniques is not going to be trivial. 
Detailed guidelines will need to be established in the SBSE/FM research community to address this important research problem that impacts all empirical studies on this subject.
\revision{To summarize the above discussion, \Cref{tab:keychallengesEmpiricalEvaluation} presents the key challenges associated with empirical evaluations.}

\begin{table}[!t]
\caption{\revision{Key challenges -- Empirical Evaluations}}
\label{tab:keychallengesEmpiricalEvaluation}
\begin{tabular}{|p{0.99\textwidth}|}
\hline
\rowcolor[HTML]{E2EDF9}
\textbf{Empirical Evaluations involving both SBSE and FMs} \\ \hline
\ding{70} Ensuring fair comparisons between FM-based and SBSE techniques, considering differences in computational resources and energy consumption.\\
\ding{70} Determining appropriate scaling factors for resource allocation to achieve fairness in experiments.\\
\ding{70} Addressing the replicability of experiments with remote, commercial FMs, which can change unpredictably over time.\\
\ding{70} Balancing the use of commercial FMs for best results with local, open-source LLMs for replicability in scientific research.\\
\ding{70} Managing non-determinism in FMs and SBSE techniques through repeated experiments and proper statistical analysis.\\
\ding{70} Establishing detailed guidelines for fair and reliable comparisons of hybrid techniques involving both FMs and SBSE.\\
\hline
\end{tabular}
\end{table}

\section{SBSE and FMs: Research Horizon for 2030}\label{sec:horizon}
In this section, we explore the future of SBSE in the context of FMs as a disruptive technology. 
We then highlight the \revision{key research opportunities} and emerging opportunities that are likely to form the research horizon for 2030.
Lastly, we discuss the challenges and opportunities of integrating SBSE and FMs in practical contexts. 

\subsection{The Future of SBSE}
\begin{figure}
\centerline{\includegraphics[width=1.0\linewidth, keepaspectratio]{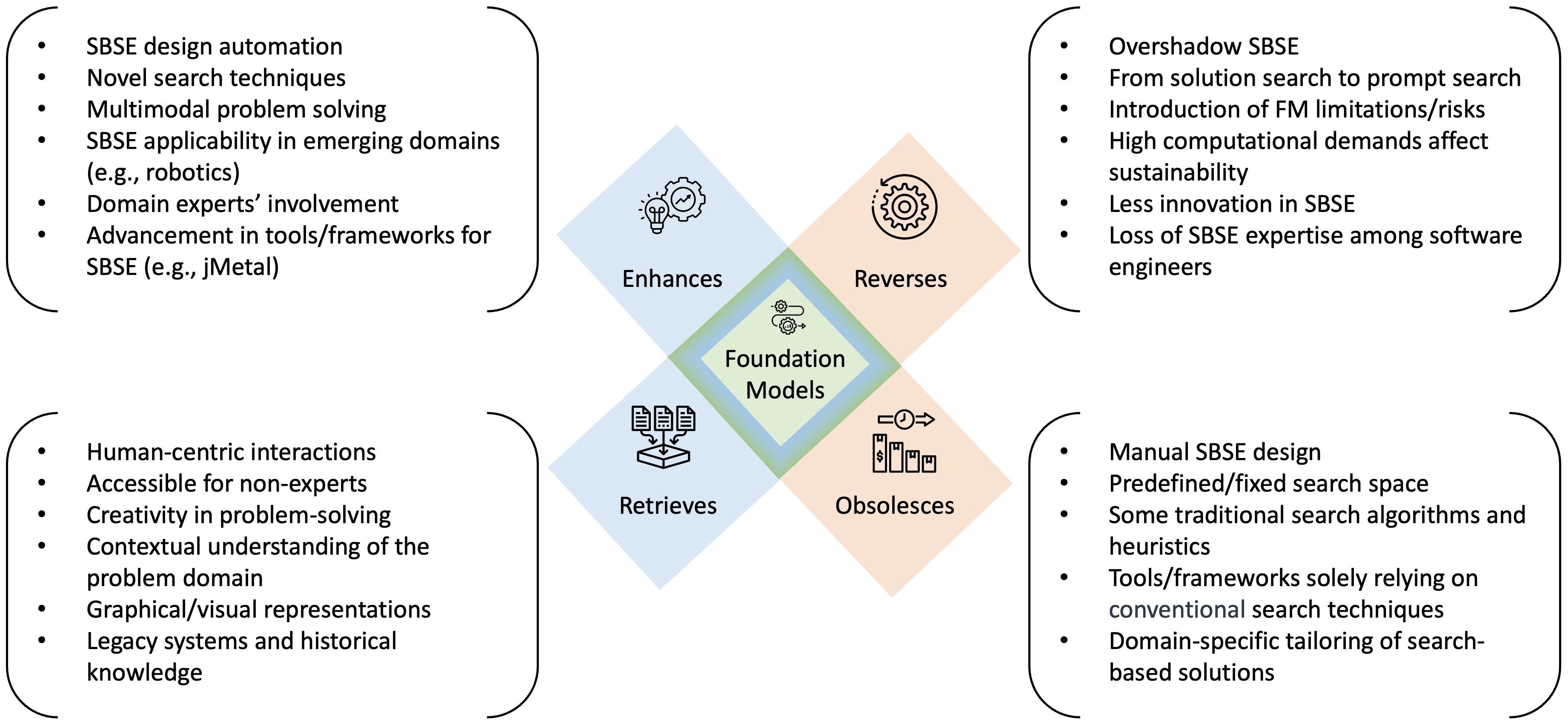}}
\caption{An overview of the tetrad illustrating the disruptive effects of FMs on SBSE.}
\label{fig:tetrad}
\end{figure}

We analyze the impact of FMs on SBSE through McLuhan’s Tetrad~\cite{mcluhan1977laws}, a framework designed to examine the disruptive effects of emerging technologies. 
The tetrad consists of four key aspects: what a new technology \textit{enhances}, \textit{retrieves}, renders \textit{obsolete}, and \textit{reverses} when pushed to extremes.
In the context of SBSE, we follow the disruptive research playbook by~\citet{storey2024disruptive} and apply McLuhan's Tetrad to assess the transformative role of FMs, considering their strengths from \fullref{FM-S1} to \fullref{FM-S4}.
Specifically, we explore four questions:
\begin{inparaenum}[(i)]
\item How do FMs \textit{enhance} SBSE?
\item What do FMs \textit{retrieve} that were previously overlooked in SBSE?
\item Which traditional SBSE techniques and tools may become \textit{obsolete} due to FMs? And
\item What does in SBSE \textit{reverse} due to over-reliance on FMs?
\end{inparaenum}
The outcomes of this analysis are summarized in~\Cref{fig:tetrad} and elaborated in the discussion that follows.
It is essential to note that, although this analysis is based on our review of the state-of-the-art, it does not provide definitive predictions of the future.

\subsubsection{Enhances}
FMs can significantly enhance automation in SBSE by reducing manual efforts across various design elements, such as population generation, solution encoding, fitness function design, crossover and mutation operators, and selection mechanisms.
This will drive the development of novel search techniques, the creation of novel FM-inspired search algorithms, and improvements over existing methodologies, such as genetic programming, by providing advanced reasoning and contextual understanding.
The multimodal capabilities of FMs will augment conventional problem-solving in SBSE, enabling the integration of diverse data types (e.g., text, images, audio, and video) and facilitating broader applicability to emerging domains such as ADS and robotics.
These capabilities will also allow domain experts to participate in the SBSE process without requiring in-depth knowledge of SBSE, fostering interdisciplinary collaboration in diverse fields, including healthcare, finance, and transportation.
Furthermore, the development of novel search techniques powered by FMs will enhance existing tools and frameworks, such as jMetal~\cite{durillo2011jmetal} and Pymoo~\cite{PymooIEEEaccess}, by integrating FM-based features like automated prompt generation, multimodal data handling, and intelligent solution evaluation.
As a result, these tools will facilitate decision-making in SBSE by providing advanced analytics and insights, such as identifying trade-offs in multi-objective optimization problems or suggesting alternative solutions, as well as customizing SBSE solutions to specific domain requirements.

\subsubsection{Retrieves}
FMs will bring back human-centric interaction to SBSE by enabling intuitive problem-solving through natural language interfaces. 
This shift allows users, even those without technical expertise, to define search problems and interpret solutions more easily. 
By reintroducing accessible, conversational workflows, FMs will make SBSE more inclusive, recalling earlier human-focused problem-solving methods that prioritized simplicity and user engagement. 
Another interesting perspective is that FMs will revive the creative and exploratory nature of problem-solving by generating novel solutions, retrieving overlooked alternatives, and providing innovative suggestions. 
Through contextual understanding, FMs will enable solutions that are based on domain-specific knowledge, reintroducing the principles of domain-driven design with enhanced scalability and intelligence. 
Moreover, FMs, such as VLMs, will reintroduce the importance of graphical and visual representations in SBSE. 
They will enhance both problem representation for various SE design artifacts, such as UML diagrams and flowcharts, and solution representation, such as visually illustrating Pareto fronts for multi-objective optimization. 
Lastly, FMs will utilize large repositories of historical data to retrieve valuable knowledge, identifying patterns in legacy code and historical project data. 
This capability will play an important role in reviving legacy systems by facilitating tasks such as understanding older codebases and enabling efficient system modernization.

\subsubsection{Reverses}
When pushed to extremes, FMs have the potential to overshadow SBSE, thus losing the significance of SBSE. 
This will shift the key focus from SBSE problem-solving mechanisms to prompt engineering as the primary means of achieving outcomes. 
As a result, the significance of SBSE, as the creative and systematic design of solutions, may be replaced by reliance on pre-trained language models. 
Over-reliance on FMs will also transfer their inherent limitations and risks, such as issues of non-determinism, hallucinations, and ethical concerns, to every SBSE application. 
Furthermore, the computational demands associated with using FMs, especially for large-scale problems, may lead to resource constraints, leading to sustainability concerns for real-world SBSE applications. 
An increasing dependence on FMs could also suppress innovation in SBSE by averting the development of novel algorithms and techniques, as the focus shifts to applying FMs rather than advancing SBSE with new methods. 
Over time, this could result in a loss of SBSE expertise among software engineers, as basic knowledge and skills are no longer exercised in practice.

\subsubsection{Obsolesces}
The integration of FMs into SBSE will render many manual design practices obsolete, such as solution encoding and fitness functions, as these tasks will become increasingly automated. 
Traditional methods that rely on predefined or fixed search spaces will also become redundant, as FMs will enable intelligent and adaptive search space definitions based on context understanding. 
Alongside these changes, several search algorithms and heuristics designed for simpler problems may become obsolete or lose relevance. 
For example, heuristics for model-based test data generation using constraint specifications (such as~\cite{ali2013generating,sartaj2025search}) will no longer be required, as FMs, particularly LLMs, can directly generate test data using these specifications.
Similarly, traditional search algorithms like the Alternating Variable Method (AVM), which were commonly used for specific SBSE problems, may be reshaped or rendered obsolete entirely. 
The obsolescence of such algorithms and heuristics will, in turn, affect the tools and frameworks that rely on them, which will need to adapt to FM-introduced changes or risk becoming irrelevant. 
Furthermore, the need for domain-specific tailoring of search-based solutions may become obsolete, as FMs provide generalization capabilities with the flexibility to adapt to a diverse range of problems and domains.

\subsection{Research Opportunities Ahead}
This section begins by summarizing the key research opportunities identified in previous sections (\Cref{sec:RootFMs4SBSE,sec:RootSBSE4FMs,sec:SBSE&FMs,sec:empiricalchallenges}) and then presents potential opportunities in emerging domains.

\subsubsection{Overview of Research Opportunities}
In this section, we present a consolidated overview of the research opportunities discussed in this roadmap by creating a mapping between the strengths and weaknesses identified in~\Cref{sec:pros&cons} and the corresponding future research directions discussed throughout the roadmap sections (\Cref{sec:RootFMs4SBSE,sec:RootSBSE4FMs,sec:SBSE&FMs,sec:empiricalchallenges}). 
Specifically, \Cref{tab:sbse_fm_mapping} summarizes how identified strengths and weaknesses motivate research opportunities across different perspectives, including using FMs to enhance SBSE design and SE artifacts generated by SBSE, using SBSE techniques to address FM limitations, and integrating SBSE and FMs to advance search processes and solve emerging SE problems.
\afterpage{%
\begin{center}
\small
\begin{longtable}{|p{3.3cm}|p{9.5cm}|p{2.5cm}|}
\caption{\revision{Mapping between the strengths/weaknesses identified in~\Cref{sec:pros&cons} and the related research opportunities discussed in the roadmap for each category from~\Cref{sec:RootFMs4SBSE,sec:RootSBSE4FMs,sec:SBSE&FMs,sec:empiricalchallenges}.}}
\label{tab:sbse_fm_mapping}\\

\hline
\rowcolor[HTML]{E2EDF9}
\textbf{Strength(s)/Weakness(es)} &
\textbf{Research Opportunity} &
\textbf{Roadmap Section}\\
\hline
\endfirsthead

\multicolumn{3}{c}%
{{\tablename\ \thetable{} -- continued from previous page}}\\
\hline
\rowcolor[HTML]{E2EDF9}
\textbf{Strength(s)/Weakness(es)} &
\textbf{Research Opportunity} &
\textbf{Roadmap Section}\\
\hline
\endhead

\hline
\multicolumn{3}{r}{{Continued on next page}}\\
\endfoot

\hline
\endlastfoot

\fullref{SB-W2}; \fullref{FM-S1}, \fullref{FM-S2} &
Automating the definition and implementation of fitness functions for SBSE problems using FMs. &
\multirow{9}{=}{FMs for SBSE Design (\Cref{ssec:roadmapfm4sbse})} \\ \cline{1-2}
\fullref{FM-S1}, \fullref{FM-S2} &
Supporting solution encoding for SBSE problems across different software engineering lifecycle phases. &
\\ \cline{1-2}
\fullref{SB-W1}; \fullref{FM-S2} &
Assisting in defining effective search spaces for SBSE problems. &
\\ \cline{1-2}
\fullref{FM-S1}, \fullref{FM-S2}, \fullref{FM-S3} &
Generating partial or complete implementations of SBSE problems using FMs. &
\\ \cline{1-2}
\fullref{FM-S2} &
Automating the testing of SBSE implementations. &
\\ \cline{1-2}
\fullref{FM-S2} &
Generating tests and test oracles for SBSE implementations. &
\\ \hline

\fullref{FM-S1}, \fullref{FM-S2} &
Interpreting and transforming SBSE-generated SE artifacts into textual and graphical representations. &
\multirow{12}{=}{FMs for SE Artifacts Generated with SBSE (\Cref{ssec:roadmapfm4seartifacts})} \\ \cline{1-2}
\fullref{FM-S2} &
Refining SBSE-generated test data using LLM-based static analysis. &
\\ \cline{1-2}
\fullref{FM-S2} &
Statically verifying programs generated with SBSE using LLMs. &
\\ \cline{1-2}
\fullref{FM-S1}, \fullref{FM-S2} &
Supporting Pareto analysis and solution selection. &
\\ \cline{1-2}
\fullref{FM-S1}, \fullref{FM-S2} &
Visualizing and intelligently pruning Pareto-optimal solutions. &
\\ \cline{1-2}
\fullref{FM-S3}, \fullref{FM-S4}; \fullref{FM-S1}, \fullref{FM-S2} &
Validating and enhancing SBSE-generated SE artifacts. &
\\ \cline{1-2}
\fullref{FM-S2} &
Improving generated test scenarios through diversity and redundancy analysis. &
\\ \cline{1-2}
\fullref{FM-S2} &
Developing LLM-based techniques for context-aware delta debugging and flaky test detection. &
\\ \hline

\fullref{FM-W2}; \fullref{SB-S3} &
Developing search-based techniques to address FM non-determinism and identify optimal software and hardware configurations. &
\multirow{11}{=}{SBSE for FM Limitations (\Cref{ssec:roadmapsbse4fms})} \\ \cline{1-2}
\fullref{FM-W2}; \fullref{SB-S1}, \fullref{SB-S2} &
Managing data-related uncertainty through search-based feature selection and FM fine-tuning. &
\\ \cline{1-2}
\fullref{FM-W2}; \fullref{SB-S1} &
Developing search-based techniques for hallucination detection and mitigation. &
\\ \cline{1-2}
\fullref{FM-W1}, \fullref{FM-W3}, \fullref{FM-W4}; \fullref{SB-S1}, \fullref{SB-S2} &
Optimizing FM fine-tuning through search-based hyperparameter and model configuration optimization. &
\\ \cline{1-2}
\fullref{FM-W2}, \fullref{FM-W5}; \fullref{SB-S1}, \fullref{SB-S2} &
Casting prompt optimization as a multi-/many-objective SBSE problem for automated prompt engineering. &
\\ \hline

\fullref{FM-W2}; \fullref{SB-S1} &
Applying search-based program improvement and code repair to refine LLM-generated code. &
\multirow{13}{=}{SBSE for SE Artifacts Generated with FMs (\Cref{ssec:roadmapfm4seartifacts})} \\ \cline{1-2}
\fullref{FM-W2}; \fullref{SB-S1} &
Applying SBST to improve test cases generated with LLMs. &
\\ \cline{1-2}
\fullref{SB-S3} &
Applying SBST for regression test selection, minimization, prioritization, and diversity of LLM-generated tests. &
\\ \cline{1-2}
\fullref{SB-S1} &
Optimizing and validating FM-generated software models using search-based techniques. &
\\ \cline{1-2}
\fullref{SB-S1}, \fullref{SB-S2} &
Optimizing FM-generated software configurations using search-based techniques. &
\\ \cline{1-2}
\fullref{SB-S1}, \fullref{SB-S2} &
Applying search-based debugging and program repair to FM-generated artifacts. &
\\ \cline{1-2}
\fullref{SB-S3} &
Evolving tests using search-based techniques based on LLM-generated fault information. &
\\ \hline

\fullref{FM-S1}, \fullref{FM-S4} &
Developing FM-inspired search algorithms with optimized initial population generation. &
\multirow{7}{=}{Advancing Metaheuristics Search in SBSE (\Cref{subsec:advsearchinsbse})} \\ \cline{1-2}
\fullref{FM-S2}, \fullref{FM-S3} &
Using FMs to calculate fitness functions during the search process. &
\\ \cline{1-2}
\fullref{FM-S2}, \fullref{FM-S3} &
Repairing invalid solutions during the search using FMs. &
\\ \cline{1-2}
\fullref{FM-S2} &
Exploring LLM-based search operators for SBSE. &
\\ \cline{1-2}
\fullref{FM-S4} &
Determining search operators dynamically at runtime using FMs. &
\\ \cline{1-2}
\fullref{FM-S4} &
Developing multimodal search operators for mixed-mode SBSE problems. &
\\ \hline

\fullref{FM-S1}, \fullref{FM-S2} &
Using LLMs to support search-based requirements engineering through fitness evaluation and Pareto analysis. &
\multirow{6}{=}{Solving SE Problems (\Cref{subsec:solvingseproblems})} \\ \cline{1-2}
\fullref{SB-S1}, \fullref{SB-S2} &
Applying SBST to identify faults in FMs for software engineering tasks. &
\\ \cline{1-2}
\fullref{SB-S1}, \fullref{SB-S2} &
Repairing FMs using search-based optimization of fine-tuning parameters. &
\\ \cline{1-2}
\fullref{SB-S1}, \fullref{SB-S2}; \fullref{FM-S1}, \fullref{FM-S4} &
Integrating SBST with multimodal FMs for GUI testing. &
\\ \hline

\fullref{SB-W1} &
Ensuring fair comparisons by using the same amount of computational resources and search budget. &
\multirow{12}{=}{Challenges of Empirical Evaluations (\Cref{sec:empiricalchallenges})} \\ \cline{1-2}
\fullref{FM-W5}, \fullref{FM-W6} &
Establishing fair resource comparisons between commercial FMs and SBSE techniques. &
\\ \cline{1-2}
\fullref{FM-W1}, \fullref{FM-W4}, \fullref{FM-W6} &
Addressing risks associated with remote commercial FMs in empirical evaluations. &
\\ \cline{1-2}
\fullref{FM-W1}, \fullref{FM-W4} &
Using open-source local FMs to improve computational cost analysis and replicability. &
\\ \cline{1-2}
\fullref{FM-W4} &
Ensuring replicability of experiments involving FMs that may change over time. &
\\ \cline{1-2}
\fullref{FM-W4} &
Establishing guidelines for fair comparisons when techniques require different hardware resources. &
\\ \hline
\end{longtable}
\end{center}
}
This mapping provides traceability between each research opportunity and the underlying capabilities and limitations of SBSE and FMs discussed in~\Cref{sec:pros&cons}.

\subsubsection{Opportunities in Emerging Domains}
FMs integrated with SBSE can be used to solve complex and dynamic nature problems. 
In the following, we present opportunities to take advantage of the synergy between SBSE and FMs in some emerging domains.

\paragraph{Autonomous Driving Systems.}
SBSE has been extensively applied to Autonomous Driving Systems (ADS), addressing various aspects of software engineering, from requirements to testing~\cite{avfuzzerISSRE20,reqsViolADSsASE21,MOSAT_FSE22,laurent2020achieving,tang2023survey,avoidCollICST2020,GambiISSTA2019,driveCharICST2021,variationsADSsTOSEM2025}. 
Since ADS rely on mixed-mode data from various sensors such as LIDAR, radar, and cameras, search-based techniques have been integrated with deep learning methods to process these datasets, particularly in simulation environments~\cite{attaoui2025search}. 
However, ADS operate in highly dynamic environments with uncertain conditions such as varying road surfaces, unpredictable weather, and fluctuating traffic patterns, which creates a significant reality gap between simulations and real-world scenarios~\cite{wu2024reality,tang2023survey,LambertenghiASE2025}. 
Therefore, considering LLMs' analytical capabilities (\fullref{FM-S2}), recent efforts have begun exploring LLMs to assist in bridging the simulation-to-reality gap~\cite{wu2025uncertainty}. 
Beyond LLMs, other FMs, such as VLMs and MMs, due to their context understanding and reasoning abilities (\fullref{FM-S4}), offer promising opportunities to enhance SBSE techniques in the ADS domain. 
One potential direction is to explore the integration of SBSE-based requirement selection and prioritization with MMs for requirement validation and traceability, enabling seamless processing of ADS requirements and multimodal sensor fusion data. 
Another direction is utilizing MMs to generate code directly from ADS requirements and design artifacts, followed by employing SBSE-driven automated program improvement techniques to optimize the generated code. 
In software testing, integrating FMs with SBSE presents opportunities such as generating realistic test scenarios for ADS, optimizing regression testing for evolving software, detecting and localizing bugs, and enhancing security testing to handle arising threats and vulnerabilities.

\revision{
These research directions have the potential to significantly improve the reliability, safety, and automation of ADS development and validation. 
From a feasibility perspective, several directions can be investigated using the current state of SBSE and FMs, whereas others will become increasingly practical as FMs, computational resources, and supporting methodologies continue to evolve. 
Given the rapid adoption of FMs in autonomous systems, these directions represent timely and high-impact opportunities for future research. 
}

\paragraph{Robotics.}
Autonomous robots employ adaptation mechanisms such as the MAPE-K loop and AI-based data analysis techniques~\cite{kephart2003vision} to autonomously monitor, analyze, and adapt to unpredictable dynamic environments and unforeseen events while interacting with humans and surrounding objects~\cite{larsen2024robotic}. 
These robots can be categorized into various types depending on the application domain, such as industrial robots for automated manufacturing, service robots for office and home assistance, and healthcare robots for medical assistance. 
Similar to ADS, these robots also process multimodal sensor and actuator data, leading to similar opportunities for integrating SBSE with FMs.

A recent shift in robotics has seen the development of robotic foundation models, including fine-tuned LLMs and VLMs for robot-specific tasks and vision action language (VLA) models designed for robots~\cite{wang2025llmforrobotics,wang2025vlmseerobotdo,din2025visionlanguageactionmodels}. 
These embodied robotic foundation models enable robots to operate in task-specific ways within closed environments while collaborating closely with humans. 
However, such interactions introduce several uncertain situations, including unpredictable human behavior, environmental changes, and task variability, raising significant reliability concerns~\cite{sartaj2025identifying,fan2025putting}. 
In this context, SBST presents promising opportunities for testing the robustness and reliability of robotic foundation models by employing FMs' capabilities, such as \fullref{FM-S2}, \fullref{FM-S3}, and \fullref{FM-S4}.  
A potential direction is to use FMs, such as VLMs and VLAs, to assist SBST in generating and simulating unpredictable human behavior, such as sudden gestures or commands, to evaluate the robot's ability to adapt and respond safely. 
Due to the black-box nature of robotic foundation models, diagnosing faults in these models is challenging~\cite{greengard2025can}. 
Therefore, another direction is to integrate FMs with SBST to detect and localize faults in robotic behaviors by analyzing multimodal data, such as correlating error logs with specific robot tasks or actions to identify inconsistencies. 
Furthermore, given the computational and time constraints associated with running extensive tests on robotic foundation models, a promising area of research lies in developing search-based test selection and prioritization techniques enhanced by multimodal FMs to focus on the most critical and impactful test cases.

\revision{
The identified research opportunities provide a promising pathway toward improving the robustness, reliability, and safety of autonomous robotic systems operating in dynamic environments. 
Several of these opportunities can be explored using existing SBSE techniques together with current robotic foundation models, particularly for simulation-based testing, fault localization, and test prioritization. Broader deployment in real-world robotic systems, however, will require further advances in embodied foundation models, multimodal reasoning, sim-to-real transfer, and validation on diverse physical platforms. 
Given the rapid evolution of robotic foundation models, including VLMs and VLAs, and their growing adoption in safety-critical applications, these directions are expected to remain a key focus of robotics research in the coming years.
}

\paragraph{AI-enabled Systems.}
Search-based approaches have been extensively used for testing~\cite{KangTOSEM2024,YooSBST2019,RiccioFSE20,ZhangTSE2022,RiccioJSHWT20} and repair~\cite{SohnTOSEM2023,distrRepICST2023,SunICSE22,NeuRecoverSANER22,adaptiveRepairGECCO2023} of ML components like deep neural networks (DNNs) used in AI-enabled systems like ADS and robotic systems, for performing tasks like perception. In test generation, a typical approach is to generate adversarial inputs; however, such inputs may be unrealistic and do not occur in practice in the real world. To tackle this issue, for image classifiers, VLMs (using capabilities \fullref{FM-S3} and \fullref{FM-S4}) could help in generating more realistic inputs by analyzing the scene and adding and/or modifying elements that could lead to misclassification. DNN repair is a popular search-based approach that modifies a few network parameters (e.g., weights) with the goal of fixing some wrong behaviors without degrading previously correct behaviors. The repair needs to be preceded by a fault localization phase~\cite{SohnTOSEM2023,DeepFaultPaper,granFaultLocDNNsISSRE2021,ma2018mode,fairFLRepTOSEM2025} that identifies the suspicious components of the network that should be repaired. While for some DNN models, like classifiers, fault localization can be quite accurate, for other models, like regression models~\cite{semSegRepSANER2025} or DNN controllers~\cite{spectacleTOSEM2025,tacticalJSS2025,dnnControllersRepairGECCO2024}, the fault localization is quite challenging. In this case, VLM (using capabilities \fullref{FM-S2} and \fullref{FM-S4}) could be used to analyze the scene and LLMs to reasons about possible causes of the violations; for example, a collision of an AI-enabled system guided by a DNN controller, could be analyzed by a VLA to identify critical frames; then, an LLM, by considering the system specification, could try to understand at which time point the controller gave a wrong control decision.

\revision{
The synergy between SBSE and foundation models offers promising opportunities to enhance testing, fault localization, and repair techniques for AI-enabled systems, ultimately improving the robustness and trustworthiness of AI components. 
From a feasibility perspective, many of these research directions can be investigated by extending existing SBST and repair frameworks with current LLMs, while more advanced capabilities, such as reasoning over complex system behaviors and multimodal execution traces, will require further methodological and technological advances. 
As AI-enabled systems are increasingly being deployed in safety-critical domains, advancing these research directions is expected to play a key role in improving the dependability of next-generation intelligent software systems. 
}

\paragraph{Internet of Things.}
Internet of Things (IoT) applications consist of heterogeneous devices, often geographically distributed, along with third-party services, all interconnected through cloud-based architectures~\cite{sartaj2023hita}. 
System-level testing of such applications with physical devices in the loop is often infeasible, making digital twins a preferred alternative for simulating these devices during testing~\cite{sartaj2024modelbased,sartaj2025medet}. 
However, search-based testing of live IoT applications, especially after each upgrade during CI/CD practices, introduces several significant challenges~\cite{sartaj2023testing,sartaj2025restapi}. 
First, running a large number of test cases after every release during CI/CD can lead to service blocking from the servers of third-party devices and service providers, potentially disrupting critical services. 
Second, fault localization becomes particularly challenging due to the interconnected nature of IoT systems. 
It is difficult to determine whether a specific failure originates from the application under test, connected devices, or third-party services. 
Third, bug reproduction is also challenging, especially for failures that occur irregularly or depend on specific runtime conditions that are difficult to replicate in testing environments. 
Fourth, identifying regressions, where previously resolved bugs reappear, or new issues arise due to updates, is another major challenge, particularly in dynamic and evolving IoT ecosystems where changes in one component can have cascading effects across the system. 
To address these challenges, more research is needed to explore the potential of integrating FMs with SBST for IoT applications. 
FMs, with their advanced analytical capabilities and contextual understanding (\fullref{FM-S1} and \fullref{FM-S2}), can enable testers to perform online system testing in a largely automated, systematized, and intelligent manner. 
For example, FMs can analyze complex interdependencies between IoT devices and third-party services to improve fault localization by identifying the root cause of failures with greater accuracy. 
In addition, FMs (using capability \fullref{FM-S1}) can enhance search-based test selection and prioritization during CI/CD by analyzing mixed-mode data from IoT devices, their digital twins, and interconnected third-party services.
For bug reproduction and regression identification, developing FM-based interactive tools to assist testers in recreating failure scenarios and identifying root causes presents a promising avenue for research. 
Furthermore, given the maturity and extensive study of search-based techniques, integrating them with recent advancements in applying LLMs for testing purposes offers another exciting direction for future exploration.

\revision{
The aforementioned research opportunities highlight how the integration of FMs and SBST can address fundamental challenges in testing complex IoT ecosystems, particularly those arising from device heterogeneity, dynamic interactions, and continuous evolution. 
In terms of feasibility, several opportunities can be explored by extending existing search-based testing frameworks with current LLMs and multimodal FMs, particularly together with digital twins and CI/CD pipelines, whereas broader deployment will require further advances in distributed reasoning, real-time analysis, and validation across heterogeneous IoT infrastructures. 
As IoT systems become increasingly interconnected and embedded in critical applications, these research directions are expected to contribute to more autonomous and effective testing approaches for future IoT ecosystems. 
}

\paragraph{Quantum Computing and SBSE.}
SBSE has also been applied to more advanced computing topics, such as quantum computing. For instance, recent works have demonstrated the use of SBSE to address quantum SE problems, including test generation~\cite{QuSBTPaper,QuSBTTool}, mutation testing~\cite{mutTGquantumProgGECCO2022}, and learning noise models for quantum computers~\cite{GP4Noise}. However, such applications remain limited, and there are ample opportunities to apply SBSE to other phases of quantum SE, such as software modeling, code generation, and test optimization. 
Moreover, current SBSE approaches primarily run on classical computers (as highlighted in \fullref{SB-W1} and \fullref{SB-W2}), although there is an increasing body of work exploring how SBSE can also be performed on quantum computers. To this end, the field of quantum search-based SE is emerging, where quantum search and optimization algorithms are applied to solve problems in classical software engineering~\cite{QAI4SE,zhao2025quantumbasedsoftwareengineering}. However, such work remains minimal, and further research is needed. Along the same lines, the application of quantum search and optimization to solve quantum SE problems (i.e., quantum search-based quantum SE) remains an almost untouched area of research. We anticipate that by 2030, more research at the intersection of quantum computing and SBSE will emerge, presenting numerous opportunities for researchers.

\revision{
The emerging intersection between quantum computing and SBSE could open new computational paradigms for solving complex software engineering problems and enable SBSE techniques to address challenges beyond the capabilities of classical computing. 
The maturity of these research directions varies, with initial studies being possible using existing quantum search and optimization algorithms, whereas large-scale applications will depend on continued progress in quantum hardware and software development. 
Given the rapid progress in quantum computing and the increasing exploration of quantum approaches for software engineering, establishing foundational knowledge at the intersection of quantum computing and SBSE will be important for shaping future research directions in this area (further details in~\cite{wang2026quantum}). 
}

\subsection{SBSE and FMs in Practice}\label{sec:industry}

At the time of writing, in 2026, there has been a major impact of LLMs on software engineering practice.  There are plenty of examples and success stories, like Microsoft's Copilot~\cite{ziegler2024measuring}, Anthropic's Claude,\footnote{https://claude.ai/}
and the ``vibe-coding'' platform Lovable.\footnote{https://lovable.dev/} 
Worldwide, startups have raised billions of euros/dollars to build products based on those technologies. 
Regardless of their actual effectiveness in solving software engineering problems, and regardless of whether all this investment in FMs is potentially leading to an economic ``bubble'',\footnote{https://www.npr.org/2025/08/30/g-s1-86377/ai-nvidia-economy-stocks} their impact on software engineering practice is undisputed.

On the other hand, comparatively, the impact of SBSE on practice has been limited. 
There are some open-source projects that have got some traction (e.g., EvoSuite,\footnote{https://github.com/EvoSuite/evosuite}
EvoMaster,\footnote{https://github.com/WebFuzzing/EvoMaster}
and 
Pynguin\footnote{https://github.com/se2p/pynguin}), and there are some successful commercialization stories
(e.g., Sapienz~\cite{mao2016sapienz,sapienz2018}).
There are also several studies done by academics with industry collaborators, where empirical studies were carried out in industry (e.g., \cite{vos2010industrial}). 
Still, this is just a ``drop-in-the-bucket'' compared to the impact of LLMs on software engineering practice. 
And, mostly, such an impact on practice has been limited to testing activities (e.g., test case generation).

There can be many different hypotheses to explain this phenomenon. 
Not only can this be related to the actual \emph{effectiveness} of the techniques at solving the problem at hand (e.g., one technique is significantly better than the other), but it can also be related to \emph{usability} concerns.
There is a difference between downloading and installing an SBSE tool, and then reading its documentation to be able to use it, compared to just typing a basic prompt in an LLM chatbox to generate test cases. 
Furthermore, \emph{awareness} and \emph{marketing} play a major role in the adoption of any technology. 
The general population can learn of the existence of LLMs just by reading the news.
But, harder to find out about Pynguin~\cite{lukasczyk2022pynguin} (for example) on your preferred news channel.

In this regard, one major selling point of FMs is their wide scope of applications. 
Writing essays from a starting prompt is one of such applications. 
And when this is used to enact cheating in university exams at large scale, that easily gets covered by newspapers worldwide.\footnote{https://www.theguardian.com/education/2025/jun/15/thousands-of-uk-university-students-caught-cheating-using-ai-artificial-intelligence-survey}
When the general population already uses FMs for several different kinds of tasks, it is not unexpected that software engineers would hear about FMs and try them out on their software engineering tasks, using the same interface (e.g., a web chat like ChatGPT\footnote{https://chatgpt.com/}).
This is the exact opposite of what happens for SBSE techniques, which are specific to only software engineering problems.
Even if there were \emph{effective} and \emph{usable} SBSE techniques that engineers might successfully use, those engineers might simply not know of their existence.
This problem is further exacerbated by academic reward systems that are more focused on academic article writing than on dissemination among practitioners.

On the one hand, when combining FMs and SBSE, there is an opportunity for SBSE to improve and strengthen the already existing impact of FMs on practice. 
On the other hand, FMs might help SBSE techniques on some specific tasks in being applicable outside the confined limits of academic research. 
Possible examples are search-based automated requirements analysis and software design.
So far, the use of SBSE to address these problems has remained confined to academic research, with apparently limited practical utility in software engineering practice. 
For highly abstract, conceptual challenges like software requirements analysis and design, automated search and optimization methods might prove ineffective. 
However, when combined with FM approaches, there might be opportunities for integrated synergies that could lead to designing more effective techniques than just using FMs alone. 
As with any complex research challenge, evaluating its feasibility and effectiveness will require future empirical studies that measure not only academic impact but also practical impact in terms of time saved, improved productivity, or adoption rates in real projects, making these approaches both more visible and more useful for practitioners.

\revision{
In summary, the integration of SBSE and FMs has the potential to bridge the gap between automated software engineering research and industrial practice by combining the broad adoption of FMs with the optimization capabilities of SBSE. 
Such integration could lead to practical solutions that improve developer productivity, automate complex software engineering tasks, and extend the applicability of SBSE beyond specialized research environments. 
Achieving this impact, however, requires addressing practical feasibility challenges, including developing user-friendly tools, integrating these approaches into existing software development workflows, and demonstrating their effectiveness through industry-driven evaluations. 
As FMs continue to rapidly transform software engineering practice, establishing synergies between FMs and SBSE represents an important near-term opportunity to ensure that SBSE research remains relevant and enable the development of more effective and practical software engineering solutions. 
}

\section{Threats to Validity}\label{sec:validitythreats}
This section presents potential threats to the validity of our proposed roadmap and the mitigation strategies used to minimize their impact. 

\paragraph{Construct Validity.}
The proposed roadmap was not developed through a formal systematic literature review. 
Although it is based on an extensive review of the relevant literature, some studies, challenges, or opportunities may have been overlooked. 
Furthermore, the identification and prioritization of research directions involve expert judgment, introducing a degree of subjectivity. 
To mitigate these threats, the roadmap was refined through structured expert discussions using the Delphi method~\cite{linstone1975delphi} and the Nominal Group Technique~\cite{harvey2012nominal}, helping improve the completeness and consistency of the identified trends and research directions. 
As a research roadmap, the objective was to identify representative rather than exhaustive research challenges and opportunities, providing a structured conceptual foundation for future research.

\paragraph{External Validity.}
The roadmap presented in this work reflects the current state of research and emerging trends at the time of writing. 
Given the rapid evolution of the field, particularly the continuous advances in foundation models (e.g., LLMs, VLMs, and VLAs), new capabilities, challenges, and research opportunities are likely to emerge beyond those identified in this work. 
Therefore, this roadmap should be viewed as a forward-looking guide rather than an exhaustive or permanent representation of the research landscape, and future updates may be required as the field continues to evolve.

\section{Related Work}\label{sec:relatedworks}
To the best of our knowledge, this paper presents the first forward-looking roadmap for SBSE in the context of FMs. 
While several existing studies have presented the state of research and future directions, many are outdated or no longer reflect the latest advancements, particularly those related to FMs.
Our work complements existing work by introducing a timely and forward-thinking vision for the evolution of SBSE in the FM era.
Initially, \citet{harman2012search} presented a detailed survey of the SBSE research state till 2010, challenges, and future directions. 
Subsequently, \citet{colanzi2020symposium} presented a systematic mapping study to analyze the evolution of SBSE from 2009 to 2019. 
\citet{sarro2023search} later presented a brief vision highlighting SBSE applications for building responsible software systems. 
Recently, \citet{wu2024evolutionary} presented a literature review and a roadmap for integrating LLMs with evolutionary algorithms (EAs). 
\revision{
The relationship between their work and ours is as follows. 
The commonalities are: 
\begin{inparaenum}[(1)]
    \item At a broader level, both works explore the synergy between foundation models and search-based optimization techniques and aim to identify future research directions.
    \item Consequently, some research opportunities and challenges, such as the development of hybrid FM-assisted optimization approaches, are common in both roadmaps.
\end{inparaenum}
The key differences are: 
\begin{inparaenum}[(1)]
    \item \citet{wu2024evolutionary} focus on integrating LLMs with EAs across various application domains, whereas our work focuses on the synergy between FMs, including, but not limited to, LLMs, and SBSE. 
    \item Although \citet{wu2024evolutionary} include software engineering as one of several EA application domains, our work focuses exclusively on SBSE, providing a comprehensive analysis of the role of FMs throughout the SBSE lifecycle. 
    \item Our work also analyzes the disruptive impact of FMs on SBSE, discussing how they reshape traditional SBSE workflows and research practices. 
    \item Moreover, we identify challenges associated with the empirical evaluation of FM-assisted SBSE techniques. 
    \item Finally, our roadmap covers research opportunities across emerging domains, considering applications of different types of FMs in addition to LLMs. 
\end{inparaenum}
}

In the context of SBST, \citet{mcminn2004search} conducted a comprehensive survey that focused on test data generation using search techniques. 
Similarly, \citet{ali2009systematic} provided a systematic review of the literature on empirical studies related to search-based test case generation up to the year 2007. 
\citet{mcminn2011search} presented a review of the SBST area, including studies till 2011, open challenges, and future directions. 
Later,~\citet{HarmanICST2015} reviewed the SBST literature up to 2014, analyzed the growth of the field, and highlighted open challenges. 
In another study, \citet{neelofar2022instance} analyzed the strengths and limitations of automated techniques in SBST, specifically focusing on instance space analysis. 
More recently, \citet{fraser2025retrospective} outlined a brief vision for the future of search-based test generation in the context of LLMs. 
In contrast, our work takes a broader perspective, exploring the synergy between SBSE and FMs (beyond just LLMs) from different aspects, challenges of empirical evaluations involving their integrated use, the disruptive effect of FMs on SBSE, and research opportunities in emerging domains.

There are some literature reviews analyzing domain-specific applications of SBSE. 
\citet{ramirez2018systematic} presented a systematic review focusing on available approaches to make an interactive search process by adding humans in the loop. 
\citet{mazzonetto2022systematic} conducted a systematic mapping study to analyze the state of research targeting SBSE for enterprise applications. 
\citet{khoshnevis2024search} presented a systematic review of SBSE applications in the software product line engineering domain. 
Compared to these works, we propose a forward-looking roadmap that highlights open challenges and research opportunities, focusing on the synergies between SBSE and FMs, their integration challenges, the potential of FMs to transform SBSE, and advancing research in emerging domains.

\section{Conclusion}\label{sec:conclusion}
Our analysis of the current state of the art revealed that most research focuses on applying large language models (LLMs) in evolutionary computation and search-based software engineering (SBSE), leaving many research opportunities to explore other FMs, like vision language models and multimodal FMs. 
We also noted some initial, yet immature, efforts in using LLMs for SBSE, SBSE for LLMs, and their combined application to solve various software engineering problems. 
Building on these insights, we presented a roadmap outlining forward-looking opportunities and research directions, focusing on the potential synergy between SBSE and FMs.
In this context, we analyzed the current research landscape, existing challenges, and promising research opportunities, organizing our analysis around three key aspects: 
\begin{inparaenum}[(1)]
\item leveraging FMs to enhance SBSE,
\item applying SBSE to advance FMs, and 
\item exploring the integration of SBSE and FMs. 
\end{inparaenum}
For empirical evaluations involving the integrated use of SBSE and FMs, we presented key challenges that require further attention from the research community. 
Furthermore, we explored the future of SBSE in the context of FMs using McLuhan's Tetrad and identified research opportunities to leverage their synergy in addressing challenges in emerging domains, covering autonomous driving systems, robotics, AI-enabled systems, Internet of Things, and quantum computing.

\section*{Acknowledgments}
This research is supported by the RoboSAPIENS project, funded through the European Commission's Horizon Europe program under Grant Agreement No. 101133807. 
Andrea Arcuri is funded by the European Research Council (ERC) under the European Union's Horizon 2020 research and innovation programme (EAST project, grant agreement No. 864972). Paolo Arcaini is supported by the ASPIRE grant No. JPMJAP2301, JST.

\bibliographystyle{plainnat}  
\bibliography{refs}  

\end{document}